%% file: ms.tex
\documentclass[times]{emulateapj}
\usepackage{lscape}

\slugcomment{Accepted for publication in ApJ}
\shorttitle{Model Comparisons}
\shortauthors{Cushing et al.}

\newcommand\teff{\mbox{$T_\mathrm{eff}$}}
\newcommand\fsed{\mbox{$f_\mathrm{sed}$}}
\newcommand\logg{\mbox{$\log g$}}
\newcommand\lbol{\mbox{$L_{\mathrm{bol}}$}}
\newcommand\fmc{\mbox{$f_\mathrm{MC}$}}

\begin{document}


\title{Atmospheric Parameters of Field L and T Dwarfs\altaffilmark{1}}


\author{Michael C. Cushing\altaffilmark{2},
        Mark S. Marley\altaffilmark{3},
        D. Saumon\altaffilmark{4},
        Brandon C. Kelly\altaffilmark{5},
        William D. Vacca \altaffilmark{6},
        John T. Rayner\altaffilmark{7},
        Richard S. Freedman\altaffilmark{8},
        Katharina Lodders\altaffilmark{9}, \&
        Thomas L. Roellig\altaffilmark{10}}

      \altaffiltext{1}{Based in part on data collected at Subaru
        Telescope, which is operated by the National Astronomical
        Observatory of Japan.}

      \altaffiltext{2}{Steward Observatory, University of Arizona, 933
        North Cherry Avenue, Tucson, AZ 85721, mcushing@ifa.hawaii.edu
        Current address: Institute of Astronomy, University of Hawai'i,
        2680 Woodlawn Drive, Honolulu, HI 96822}

      \altaffiltext{3}{NASA Ames Research Center, MS 254-3, Moffett
        Field, CA 94035, Mark.S.Marley@NASA.gov}

      \altaffiltext{4}{Los Alamos National Laboratory, Applied Physics
        Division, MS F663, Los Alamos, NM 87545, dsaumon@lanl.gov}

      \altaffiltext{5}{Steward Observatory, University of Arizona, 933
        North Cherry Avenue, Tucson, AZ 85721,
        bkelly@as.arizona.edu}

      \altaffiltext{6}{SOFIA-USRA, NASA Ames Research Center MS N211-3,
        Moffett Field, CA 94035, wvacca@mail.arc.nasa.gov}

      \altaffiltext{7}{Institute for Astronomy, University of Hawai`i,
        2680 Woodlawn Drive, Honolulu, HI 96822,
        rayner@ifa.hawaii.edu}

      \altaffiltext{8}{SETI Institute, NASA Ames Research Center, MS
        254-5, Moffett Field, CA 94035, freedman@darkstar.arc.nasa.gov}

      \altaffiltext{9}{Planetary Chemistry Laboratory, Department of
        Earth and Planetary Sciences, Washington University, St.
        Louis, MO, 63130, lodders@levee.wustl.edu}

      \altaffiltext{10}{NASA Ames Research Center MS N245-6, Moffett
        Field, CA 94035, thomas.l.roellig@nasa.gov}

\begin{abstract}

  We present an analysis of the 0.95$-$14.5 $\mu$m spectral energy
  distributions of nine field ultracool dwarfs with spectral types
  ranging from L1 to T4.5.  Effective temperatures, gravities, and
  condensate cloud sedimentation efficiencies are derived by comparing
  the data to synthetic spectra computed from atmospheric models that
  self-consistently include the formation of condensate clouds.  Overall
  the model spectra fit the data well, although the agreement at some
  wavelengths remains poor due to remaining inadequacies in the models.
  Derived effective temperatures decrease steadily through the L1 to
  T4.5 spectral types and we confirm that the effective temperatures of
  ultracool dwarfs at the L/T transition are nearly constant, decreasing
  by only $\sim$200 K from spectral types L7.5 to T4.5.  The condensate
  cloud properties vary significantly among the L dwarfs in our sample,
  ranging from very thick clouds to relatively thin clouds with no
  particular trend with spectral type.  The two objects in our sample
  with very red $J-K_s$ colors are, however, best fitted with synthetic
  spectra that have thick clouds which hints at a possible correlation
  between the near-infrared colors of L dwarfs and the condensate cloud
  properties.  The fits to the two T dwarfs in our sample (T2 and T4.5)
  also suggest that the clouds become thinner in this spectral class, in
  agreement with previous studies.  Restricting the fits to narrower
  wavelength ranges (i.e., individual photometric bands) almost always
  yields excellent agreement between the data and models.  In some
  cases, the models even reproduce the detailed structure of the weak
  H$_2$O and CH$_4$ absorption features.  Limitations in our knowledge
  of the opacities of key absorbers such as FeH, VO, and CH$_4$ at
  certain wavelengths remain obvious, however.  The effective
  temperatures obtained by fitting the narrower wavelength ranges can
  show a large scatter compared to the values derived by fitting the
  full spectral energy distributions; deviations are typically $\sim$200
  K and in the worst cases, up to 700 K.  In spite of the uncertainties
  remaining in the models, the approach presented here works rather well
  and should be extended to a much larger sample.

\end{abstract}

\section{Introduction} \label{sec:Introduction}

The emergent spectra of very low-mass (VLM) stars and brown dwarfs,
collectively known as ultracool dwarfs, are shaped in large part by gas
and condensation chemistry.  The low temperature (500 K $\lesssim$ $T$
$\lesssim$ 3000 K), high pressure (1 bar $\lesssim$ $P$ $\lesssim$ 10
bar) conditions typical of ultracool dwarf atmospheres favor the
formation of molecules (such as CO, CH$_4$, N$_2$, NH$_3$, and H$_2$O)
and, at $T$ $<$ 2400 K condensates from the refractory elements (Ti, V,
Ca, Al, Fe, Si, and Mg).  Condensation removes gaseous opacity sources
(such as TiO, VO, FeH, CrH) from the atmosphere and alters the
atmospheric chemistry as the condensates gravitationally settle and form
clouds that deplete the upper layers of the atmosphere of both the
condensates and their constituent elements
\citep{1994Icar..110..117F,1996ApJ...472L..37F,1999ApJ...519..793L,
  2002ApJ...577..974L,1999ApJ...512..843B}.  Condensates also contribute
their own opacity, which can be substantial for the more abundant Fe-,
Mg-, and Si-bearing species such as forsterite (Mg$_2$SiO$_4$),
enstatite (MgSiO$_3$), and solid or liquid Fe
\citep[e.g.,][]{2002ApJ...568..335M,2002ApJ...575..264T}.

This rich chemistry gives rise to complex emergent spectra that are
distinctly non-Planckian.  The molecular line opacities dominate over
all of the continuum opacity sources (e.g., H$^-$) that are important in
the atmospheres of hotter stars \citep{1995ApJ...445..433A} and as a
result the emergent spectra of ultracool dwarfs exhibit strong molecular
absorption bands of TiO, H$_2$O, and CH$_4$.  Indeed, it is the strength
of these and other molecular bands that are used to define the L and T
spectral classes \citep{1999ApJ...519..802K,2006ApJ...637.1067B}.  L
dwarf (2400 K $\lesssim$ \teff\ $\lesssim$ 1400 K) spectra exhibit
weakening TiO and VO bands at red optical wavelengths (0.6$-$1.0 $\mu$m)
due to the formation of Ti-bearing condensates such as perovskite
(CaTiO$_3$).  The loss of the gaseous TiO and VO opacity results in a
more transparent atmosphere and leads to the emergence of monatomic
alkali lines (Na, K, Rb, Cs).  Most prominent are the resonance
\ion{Na}{1} (5890/5896 \AA) and \ion{K}{1} (7665/7699 \AA) doublets,
which become increasingly broad with later spectral type.  At
near-infrared wavelengths (1$-$2.5 $\mu$m) the L dwarfs also become
progressively redder with later spectral types due to the additional
condensate opacity.  As CH$_4$ becomes the dominant carbon-bearing gas
in the upper, coolest layers of the atmosphere (CO/CH$_4 <$ 1 for
$T\lesssim$ 1100 K at $P$=1 bar), CH$_4$ absorption bands emerge at
near-infrared wavelengths which signals the transition between the L and
T (1400 K $\lesssim$ \teff\ $\lesssim$ 700 K) spectral classes.  The
\ion{K}{1} and \ion{Na}{1} resonance lines are present in the red
optical spectra of T dwarfs \citep{2003ApJ...594..510B} because the high
temperature condensate anorthite (CaAl$_2$Si$_2$O$_8$), a feldspar, has
gravitationally settled out of the cooler atmospheric regions.  In the
absence of such settling, Na and K would condense into a feldspar in the
form of albite (NaAlSi$_3$O$_8$) and orthoclase (KAlSi$_3$O$_8$)
\citep{1999ApJ...512..843B,1999ApJ...519..793L,2002ApJ...568..335M}.

The observed spectra and photometry of ultracool dwarfs can therefore
serve as touchstones against which we can test our understanding of
atmospheric physics, chemistry, dynamics, and cloud processes as
embodied in our atmospheric models.  Shortly after the discovery of the
first bona fide brown dwarf, Gliese 229B \citep{1995Natur.378..463N}, a
number of publications appeared comparing the observed spectrum to
models
\citep{1996ApJ...465L.123A,1996Sci...272.1919M,1996A&A...308L..29T}.
\citet{2001RvMP...73..719B} presented a review of the substantial
subsequent work devoted to understanding the properties of L and T
dwarfs.  Although the comparison of models to the spectra of individual
objects such as Gl 229B \citep{2000ApJ...541..374S} and Gl 570D
\citep{2001ApJ...556..373G,2006ApJ...647..552S} has played an important
role in the progression of the field, relatively little work has been
done on comparing entire classes of models to uniform datasets.  Such
comparisons are important as they can reveal the global properties of a
set of objects and highlight the systematic strengths and weaknesses of
the models.

In this work, we present such a comparison to one particular set of
atmospheric models \citep[][2007, in preparation]{2002ApJ...568..335M}.
We aim to understand what the modeling has accomplished to date and to
help direct future models and observations along the pathways that seem
most promising.  The paper is organized as follows.  Sections \S2 and
\S3 present the data and model atmospheres used in the analysis,
respectively.  In \S4, we present the results of the comparisons between
the data and models and in \S\ref{sec:Systematic Uncertainties}, we
discuss two systematic uncertainties that affect our analysis.

\section{The Data} \label{sec:The Data}

There are currently eleven L and T dwarfs that have published spectra
with nearly complete wavelength coverage from 0.65 to 14.5 $\mu$m
\citep{2006ApJ...648..614C,2007ApJ...656.1136S}.  We selected the 5 L
and 2 T dwarfs from the \citeauthor{2006ApJ...648..614C} sample with the
highest S/N spectra that are not known to be multiple systems.  We also
included two additional L dwarfs, 2MASS J1506514$+$1321060 and DENIS
J025503.3$-$470049.0, that were not part of the
\citeauthor{2006ApJ...648..614C} sample.  The properties of the dwarfs
in our sample, including optical and infrared spectral types and Two
Micron All Sky Survey \citep[2MASS; ][]{2006AJ....131.1163S} photometry,
are given in Table \ref{tab:objlist}.  Although both optical and
infrared spectral types are listed in Table \ref{tab:objlist}, we
hereafter use optical types for the L dwarfs \citep{1999ApJ...519..802K}
and infrared types for the T dwarfs \citep{2006ApJ...637.1067B}.  In
addition, we abbreviate the numerical portions of the 2MASS, Sloan
Digital Sky Survey \citep[SDSS;][]{2000AJ....120.1579Y}, and Deep Near
Infrared Survey of the Southern Sky
\citep[DENIS;][]{1997Msngr..87...27E} target designations as
hhmm$\pm$ddmm, where the suffix is the sexigesimal Right Ascension
(hours and minutes) and declination (degrees and arcminutes) at J2000
equinox.

The 0.65$-$14.5 $\mu$m spectra used in this work were constructed from
the following sources: red-optical spectra were obtained with the Low
Resolution Imaging Spectrograph \citep[LRIS; ][]{1995PASP..107..375O} at
the Keck telescope; near-infrared spectra were obtained with SpeX
\citep{2003PASP..115..362R} at the NASA IRTF and the Infrared Camera and
Spectrograph \citep[IRCS; ][]{2000SPIE.4008.1056K} on Subaru; and
mid-infrared (5.5$-$14.5 $\mu$m) spectra were obtained with the Infrared
Spectrograph \citep[IRS; ][]{2004ApJS..154...18H} on board the
\textit{Spitzer Space Telescope} \citep{2004ApJS..154....1W}.  The only
wavelengths not covered by these spectra are those containing the strong
telluric absorption bands at 1.9 and 2.6 $\mu$m and the 4.5$-$5.5 $\mu$m
range.  The latter wavelength range is exceedingly difficult to observe
from the ground and is not covered by the IRS.  Table \ref{tab:specinfo}
lists both the references and resolving powers of the various spectra
used in this work.  For each dwarf, the different spectra were
absolutely flux calibrated using ground-based photometry and then
combined into a single 0.65$-$14.5 $\mu$m spectrum as described in
\citet{2005ApJ...623.1115C,2006ApJ...648..614C}.

\begin{figure} 
\centerline{\hbox{\includegraphics[scale=1.5]{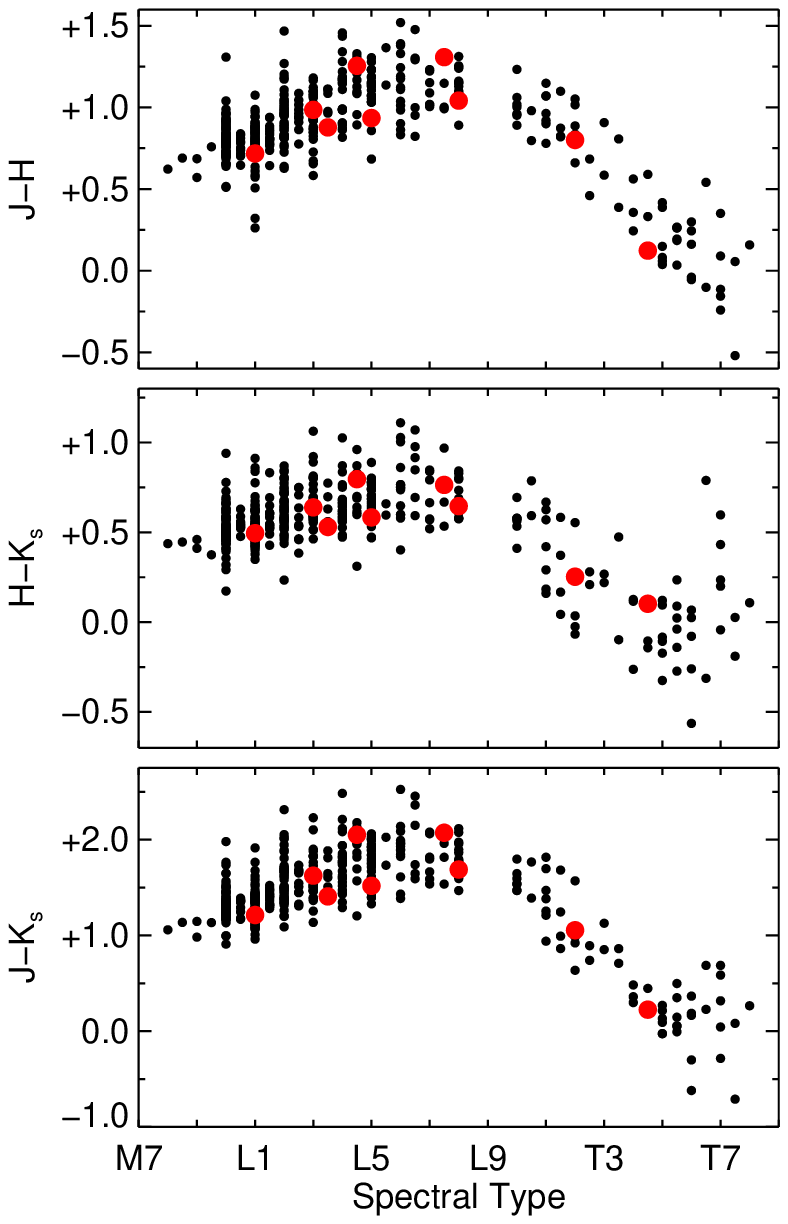}}}
\caption{\label{fig:colors}2MASS $J-H$ (\textit{top}), $H-K_S$
  (\textit{middle}), and $J-K_s$ (\textit{bottom}) colors as a function
  of spectral type for all of the L and T dwarfs in the DwarfArchives
  web site as of 2007$-$06$-$07.  The dwarfs in our sample are shown in
  red.}
\end{figure}

Figure \ref{fig:colors} shows the 2MASS $J-H$, $H-K_s$, and $J-K_s$
colors as a function of spectral type for all of the L and T dwarfs
listed in the Dwarf Archives\footnote{\url{http://DwarfArchives.org}}
web site as of 2007 June 07.  The dwarfs in our sample are shown as red
points.  Overall, our sample is representative of dwarfs with similar
spectral types.  However, 2MASS 2224$-$0158 (L4.5) and 2MASS 0825$+$2115
(L7.5) have relatively red $J-K_s$ colors given their spectral types and
2MASS J1507$-$1627 (L5) has a relatively blue $J-K_s$ color.  We discuss
these dwarfs in more detail in \S\ref{sec:Very Red L Dwarfs}.

\section{The Models} \label{sec:The Models}

The L and T dwarf models presented here have been produced in the same
way as those in our earlier work \citep[e.g.,
][]{1996Sci...272.1919M,1997ApJ...491..856B,2002ApJ...568..335M,
  2006ApJ...647..552S}.  We are preparing a complete description of the
models for a future publication.  Briefly, we computed a grid of
one-dimensional, plane-parallel, hydrostatic, non-gray,
radiative-convective equilibrium atmosphere models using our standard
modeling code.  A one-dimensional, plane parallel approximation is
appropriate since the atmospheric scale height is always much smaller
than the ultracool dwarf radius.  We do not use mixing length theory,
but rather set the local lapse rate in the convective regions to the
adiabatic lapse rate.  This is an excellent approximation in the dense,
warm atmospheres of brown dwarfs and giant planets where convection is a
very efficient energy transport mechanism.  The superadiabaticity is
very small and the atmospheric profile is insensitive to the choice of
mixing length \citep{1997A&A...327.1054B, 1997ARA&A..35..137A}.

For the chemical equilibrium calculations we use the elemental abundance
data of \citet{2003ApJ...591.1220L} and compute compositions following
\citet{1994Icar..110..117F}, \cite{2002Icar..155..393L}, and
\citet{1999ApJ...519..793L,2002ApJ...577..974L}.  The chemical
equilibrium abundances are computed for local thermodynamic equilibrium
conditions.  In stellar atmospheres, departures from thermochemical
equilibrium can arise from interactions of atoms and molecules with the
non-thermal radiation field while in planetary atmospheres,
non-equilibrium conditions can arise from vertical transport (by
convection or eddy diffusion) on a time scale shorter than that of the
chemical reactions involved.  In ultracool dwarf atmospheres, the first
effect is negligible compared to the second.  In the present work, we
neglect both effects.  The opacity database includes the molecular lines
of H$_2$O, CH$_4$, CO, NH$_3$, H$_2$S, PH$_3$, TiO, VO, CrH, FeH,
CO$_2$, HCN, C$_2$H$_2$, C$_2$H$_4$, and C$_2$H$_6$, the atomic lines of
the alkali metals (Li, Na, K, Rb and Cs), and continuum opacity from
H$_2$ collision induced absorption (CIA), Rayleigh scattering from
H$_2$, H and He, bound-free opacity from H, H$^-$, and H$_2^+$, and
free-free opacity from He, H$_2^-$, and H$_2^+$.  The molecular line
lists are continually updated and while they have well-known
shortcomings, they represent the state-of-the-art.  An explanation of
the treatment of line broadening, and further details regarding the
chemical equilibrium calculation are presented in \citet{freedman07}.
Note that we do not include a turbulent velocity in the calculation of
line shapes as test calculations have shown this has a negligible effect
for the temperatures and pressures considered here.  An error in the TiO
line opacity database \citep{1998cpmg.conf..321S} employed in our recent
generations of models (including those in \citet{2002ApJ...568..335M},
\citet{2002ApJ...571L.151B}, and \citet{2004AJ....127.3553K}) that
resulted in too-strong TiO absorption has been corrected.

For the model atmosphere calculation we use the $k$-coefficient method
to describe the opacities \citep{1989JQSRT..43..191G}.  Within each of
180 spectral bins we sum the opacity arising from {\it each} of the
several hundred million atomic and molecular lines in our database.  Our
radiative transfer follows the source function technique
\citep{1989JGR....9416287T}, allowing inclusion of arbitrary Mie
scattering particles in the layer opacities.  This approach is exact in
the limit of single scattering and treats multiple scattering by
assuming that the scattered radiation field can be approximated with two
streams.  After computing a pressure-temperature ($P,T$) profile with
the $k$-coefficient method for a given set of parameters, we compute a
high resolution spectrum by solving the monochromatic radiative transfer
equation for a larger number of frequency points (typically $\sim
190000$ in the range of 0.4 to 50$\,\mu$m), using the exact same
molecular line lists, continuum opacity sources, chemical composition,
and line profiles as those used for the computation of the
$k$-coefficients.  The resulting spectra can be smoothed or binned for
comparison with data.  Previous applications of our atmospheres code
include the generation of $P$/$T$ profiles and spectra for Titan
\citep{1989Icar...80...23M}, Uranus \citep{1999Icar..138..268M}, hot
Jupiters \citep{2005ApJ...627L..69F,2006ApJ...642..495F}, and brown
dwarfs \citep{1996Sci...272.1919M,1997ApJ...491..856B,
  2002ApJ...568..335M, 2006ApJ...647..552S}.
  
As we converge to a model temperature-pressure profile, we compute the
condensate size and vertical profiles following the prescription of
\citet{2001ApJ...556..872A} for various values of the cloud
sedimentation efficiency parameter, \fsed.  Larger values of \fsed\
imply larger particle sizes and thus greater sedimentation efficiency.
The width of the lognormal particle size distribution is set at $\sigma
= 2$ and the eddy diffusion coefficient above the convective region is
set to $K_{\rm zz}= 10^{5}\,\rm cm^2\,sec^{-1}$.  This latter value is
consistent with that found from studies of the effect of non-equilibrium
chemistry in L and T dwarf atmospheres
\citep{2006ApJ...647..552S,2007ApJ...655.1079L}.  For any $P/T$ profile
there is a single description of the variation in particle sizes and
number densities with altitude for each of the condensed species.  As
the atmospheric structure is iterated towards the equilibrium solution
the cloud description is also continuously updated.  The final result is
a single, self-consistent thermal and cloud profile with the specified
gravity, effective temperature, and cloud sedimentation efficiency.  We
term these models `cloudy'.  We also consider models in which the
condensate formation is included in the chemical equilibrium
calculation, but the condensate opacity is neglected.  We term these
models `clear' and denote them as \fsed=nc, since the models include 'no
cloud' opacity.

In this first study, for the sake of simplicity, we consider a model
parameter space consisting of \teff, $g$, and \fsed\ at solar
metallicity.  A more extensive study including variations in metallicity
and the other cloud parameters, as well as departures from chemical
equilibrium caused by vertical mixing in the atmosphere
(e.g. \citet{2006ApJ...647..552S}), will be the subject of future
papers.  Evolutionary models predict that field L and T dwarfs with ages
greater than a few hundred million years have 700 K $\lesssim$ \teff\
$\lesssim$ 2400 K and 5.0 $\lesssim$ \logg\ (cm s$^{-2}$) $\lesssim$ 5.5
\citep{1997ApJ...491..856B}.  We consider the entire range of \teff\ in
steps of 100 K, and values of \logg\ of 4.5, 5.0, and 5.5. In previous
work found that \fsed\ $\sim 3$ best reproduces global properties of
Jupiter's ammonia cloud \citep{2001ApJ...556..872A} and the $J-K$ colors
of L dwarfs \citep[e.g., ][]{2004AJ....127.3553K}.  However the excess
TiO opacity in our previous models, particularly in the $J$ band, may
have led us to overestimate the \fsed\ value for L dwarfs.  Accordingly
here we consider models with \fsed=1, 2, 3, 4, and nc.  The final grid
contains 270 models.

Dynamical processes in the atmospheres of individual objects ultimately
control the growth and sedimentation of condensates and the optical
properties of the various cloud decks expected in these atmospheres
\citep[e.g.,][]{2001A&A...376..194H,2003A&A...399..297W,2004A&A...414..335W,2004A&A...423..657H,2006A&A...455..325H}.
Although our modeling parameter, \fsed\ aims to self-consistently
capture these processes in a tractable way, the global properties of
each individual object must ultimately define the true properties of the
clouds.  However we do not yet understand how differences in
metallicity, rotation rate, effective temperature, and gravity might
manifest themselves as different values of \fsed.  Therefore while we
treat this parameter on equal footing as $g$ and \teff\ for fitting
purposes, we recognize that ultimately, more complete models would find
that \fsed\ is determined by the global physical properties of
individual objects.

\begin{figure} 
\centerline{\hbox{\includegraphics[scale=1.5]{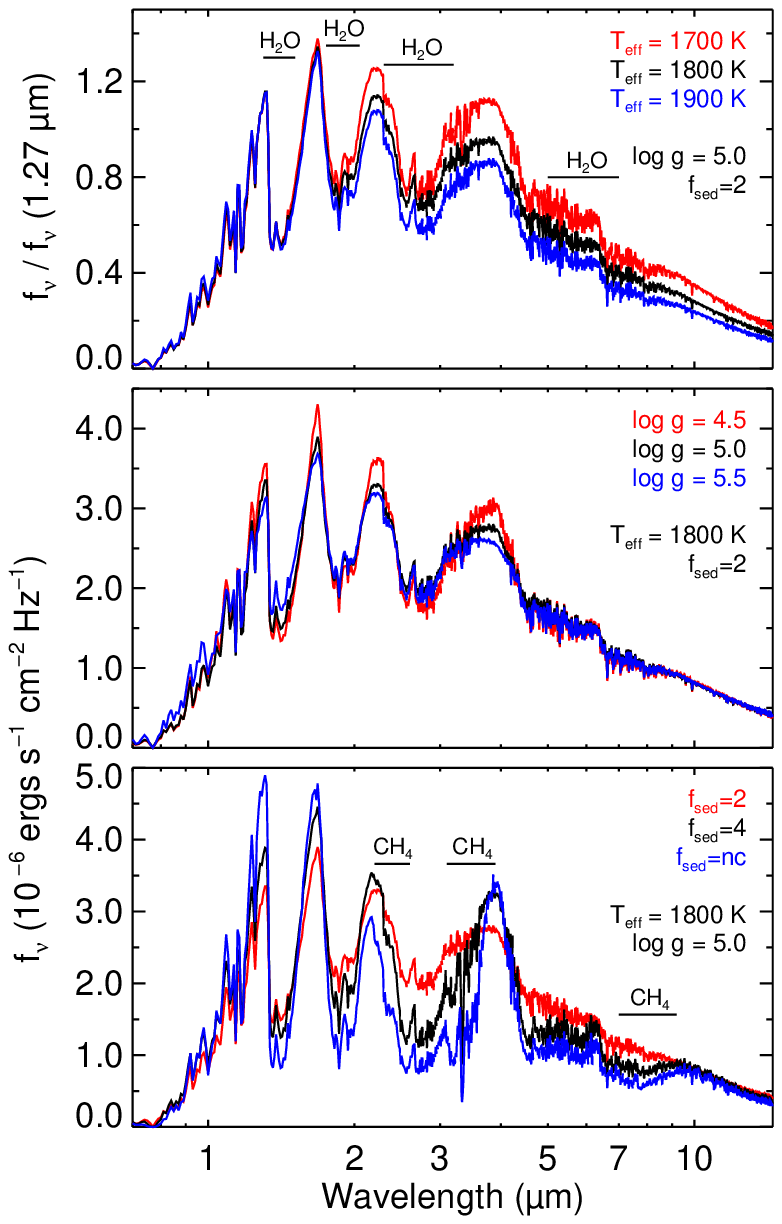}}}
\caption{\label{fig:1800KVariations}A comparison of synthetic spectra
  with atmospheric parameters typical of an L dwarf (\teff=1800 K,
  \logg=5.0 cm s$^{-2}$, \fsed=2) when \teff, \logg, and \fsed\ are
  varied.  The models have been smoothed to $R$=500 and normalized to
  unity at 1.27 $\mu$m in the top panel to show the relative variations
  in the models as a function of \teff.  The flux density units of the
  models in the lower two panels correspond to the emergent flux at the
  top of the atmosphere.}
\end{figure}

\begin{figure} 
\centerline{\hbox{\includegraphics[scale=1.5]{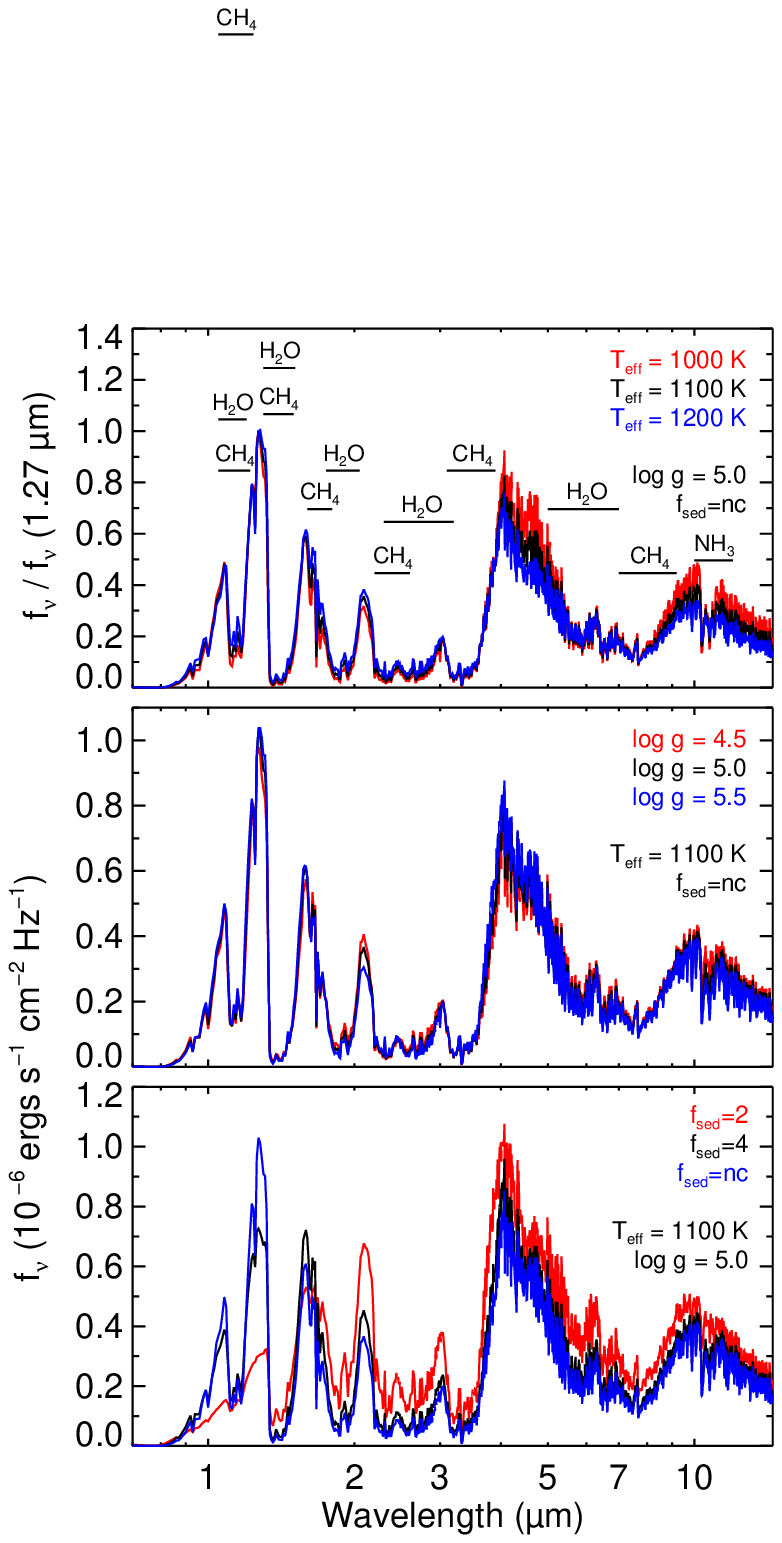}}}
\caption{\label{fig:1100KVariations}Same as Figure
  \ref{fig:1800KVariations} except for the atmospheric parameters of a
  typical T dwarf (\teff=1100 K, \logg=5.0 cm s$^{-2}$, \fsed=nc).}
\end{figure}

For each combination of gravity, effective temperature, and
sedimentation efficiency, we compute a self-consistent atmospheric
temperature-pressure profile.  Using this profile and employing the same
atmospheric chemistry and opacities, we compute a high resolution
synthetic spectrum.  This spectrum is then smoothed to the resolving
power of the data using a Gaussian kernel and resampled onto the same
wavelength grid.  Figures \ref{fig:1800KVariations} and
\ref{fig:1100KVariations} show a sequence of synthetic spectra, smoothed
to $R$=500 for clarity, with atmospheric parameters typical of an L
dwarf (\teff=1800 K, \logg=5.0, \fsed=2) and a T dwarf (\teff=1100 K,
\logg=5.0, \fsed=nc).  The three panels in each figure illustrate the
variations in spectral morphology when just one of the three parameters
is changed.

The upper panels show the changes in spectral morphology due to
variations in \teff\ at a fixed $g$ and \fsed.  With decreasing \teff,
the L dwarfs (Fig. \ref{fig:1800KVariations}) become redder in the
near-infrared and the depths of the H$_2$O bands become deeper.  Note
also that the Q branch of the $\nu_3$ fundamental band of CH$_4$ at 3.3
$\mu$m is present at \teff=1700 K.  The depths of the H$_2$O, CH$_4$,
and NH$_3$ bands in the spectra of the T dwarfs
(Fig. \ref{fig:1100KVariations}) become deeper with decreasing \teff\
but in contrast to the L dwarfs, the T dwarfs become slightly bluer in
the near-infrared.

The middle panels show the changes in spectral morphology due to
variations in $g$ at a fixed \teff\ and \fsed. In the L dwarfs the
primary effect of increasing $g$ is to weaken the near-infrared H$_2$O
absorption bands.  At these temperatures, the molecular abundance of
H$_2$O and CH$_4$ depend weakly on the atmospheric pressure and are thus
not sensitive to changes in gravity.  The change in the H$_2$O band
strengths is instead a result of the fact that, at a given pressure
level in the atmosphere, the higher gravity models are cooler and as a
result have thicker condensate clouds.  The clouds produce a grayer
emergent spectrum and are therefore responsible for the decreased depth
of the H$_2$O bands.  In contrast to the L dwarfs, the H$_2$O and CH$_4$
band depths in the spectra of T dwarfs are only marginally affected by
changes in $g$, except in the $K$ band.  The peak flux level in the $K$
band diminishes with increasing $g$ due to an increase in the CIA of
H$_2$.  This broad, featureless absorption band
\citep[e.g.,][]{1997A&A...324..185B} centered at 2.4 $\mu$m is very
sensitive to atmospheric pressure, and thus $g$, since it involves the
collisions of particles ($\kappa_\nu \propto n_{gas}^2$, where $n_{gas}$
is the number density of the gas).

Finally, the lower panels show the changes in spectral morphology due to
variations in the cloud sedimentation parameter \fsed\ at a fixed \teff\
and $g$.  Changes in \fsed\ produce the most dramatic variations in
spectral morphology (particularly in the 1$-$6 $\mu$m region)
underscoring the importance of correctly modeling the formation and
subsequent sedimentation of condensates.  The additional condensate
opacity results in emergent spectra that are redder at near-infrared
wavelengths than those from the cloudless models (\fsed=nc).  The impact
of the clouds on the emergent spectra of ultracool dwarfs becomes more
dramatic with lower \fsed\ values because the clouds are thicker and
thus contribute more opacity to the atmosphere.

\section{Atmospheric  Parameters of L and T Dwarfs}

\subsection{Fitting Technique} \label{sec:The Fitting Technique}

The 1D model atmospheres predict the emergent flux density
$\mathcal{F}_\nu$ ($\mathcal{F}_\nu = 2\pi\int_0^\infty I_\nu\mu\,d\mu$)
as a function of wavelength (i.e., a synthetic spectrum) at the surface
of a star.  The synthetic spectra must be multiplied by $(R/d)^2$, where
$R$ is the stellar radius and $d$ is the stellar distance, in order to
obtain fluxes at Earth and directly compare them to observations.
Although the parallaxes of nearly 80 L and T dwarfs have been measured
\citep[e.g,
][]{2002AJ....124.1170D,2003AJ....126..975T,2004AJ....127.2948V,
  2006AJ....132.1234C}, the radii of field L and T dwarfs are currently
unknown.  Therefore we can compare only the relative shapes of the
modeled spectral energy distributions (SEDs) to the data.

For each model $k$, which hereafter is denoted by the triplet
[\teff/\logg/\fsed], we compute a goodness-of-fit statistic $G_k$
defined by\footnote{We also tested the usefulness of the statistic $G_k
  = \sum_{i=1}^n w_i \left ( \frac{f_i -
      C\mathcal{F}_{k,i}}{f_i}\right)^2$ since most published spectra do
  not have errors associated with the flux density values.  However as
  noted by \citet{1995PASJ...47..287T}, such a statistic gives more
  weight to wavelengths with low flux values.  Typically ultracool dwarf
  spectra have very low S/N at such wavelengths and thus care must be
  taken when using such a statistic.},

\begin{equation}
\label{eq:G}
G_k = \sum_{i=1}^n w_i \left ( \frac{f_i - C_k\mathcal{F}_{k,i}}{\sigma_i} 
\right)^2,
\end{equation}
\noindent
where $n$ is the number of data pixels, $w_i$ is the weight for the
$i$th wavelength, $f_i$ and $\mathcal{F}_{k,i}$ are the flux densities
of the data and model $k$, respectively, $\sigma_i$ are the errors in
the observed flux densities, and $C_k$ is the unknown multiplicative
constant equal to ($R/d$)$^2$.  For each model $k$, the constant $C_k$
is determined by minimizing $G_k$ with respect to $C_k$ and is given by,

\begin{equation}
\label{eq:C}
C_k = \frac{\sum w_i f_i \mathcal{F}_{k,i} / \sigma_i^2}{\sum w_i 
  \mathcal{F}_{k,i}^2 / \sigma_i^2}.
\end{equation}
\noindent
We select the best fitting synthetic spectrum by locating the global
minimum of the $G$ values for all of the model spectra in our grid.

We have no \textit{a priori} reason to favor one wavelength range over
another and therefore, in principle, all wavelengths points could
receive equal weight in our fits.  However given the large variation in
the wavelength sampling of our data (e.g, the near-infrared spectra
cover only $\sim$2 $\mu$m but have $\sim$20 times more wavelength points
than the \textit{Spitzer} IRS spectra which cover $\sim$10 $\mu$m), $G$
would be heavily biased towards near-infrared wavelengths if $w_i$ was
set to unity for all $i$.  We therefore chose to weight each pixel by
its width in microns ($w_i$ = $\Delta \lambda_i$).  In
\S\ref{sec:Weights}, we discuss the systematic effects that our choice
of weights has on our results.  Since the line lists of CH$_4$ are
incomplete in the $H$-band \citep{2001RvMP...73..719B,freedman07}, and
the $E$ $^4\Pi$ $-$ $A$ $^4\Pi$ system of FeH at 1.6 $\mu$m which is
present in the spectra of the L dwarfs
\citep{2001ApJ...559..424W,2003ApJ...582.1066C} lacks a line list, we
also set the weights of the pixels from 1.58 to 1.75 $\mu$m (roughly the
center of the $H$ band) to zero.

\begin{figure} 
  \centerline{\hbox{\includegraphics[scale=1.5]{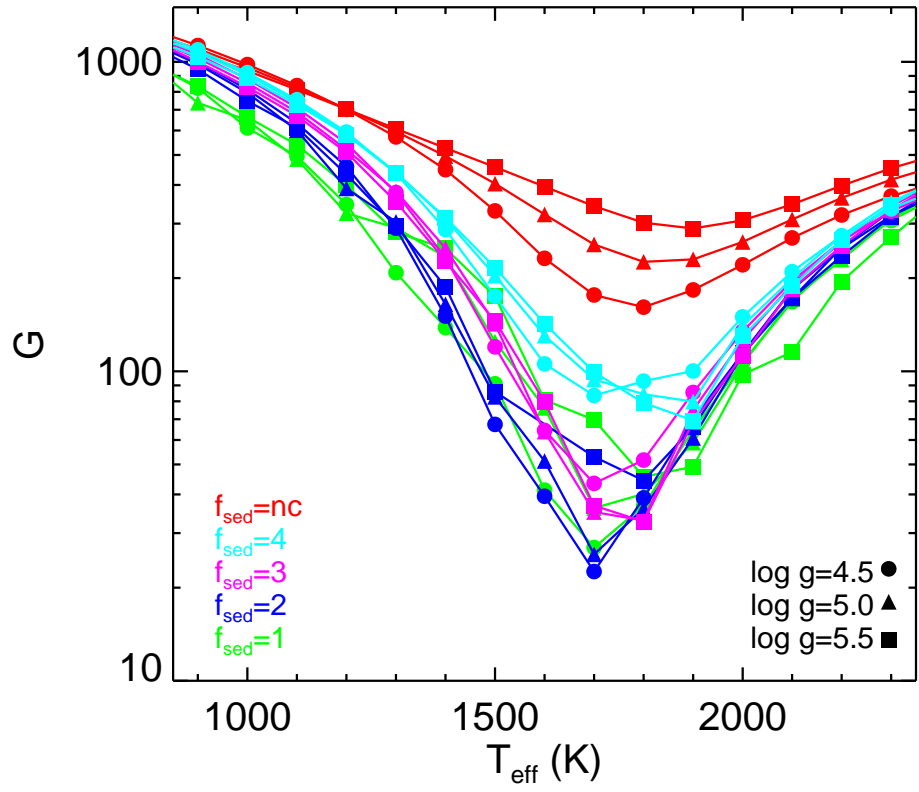}}}
  \caption{\label{fig:Chi2}The $G$ values as a function of \teff\ for
    the 0.95$-$14.5 $\mu$m spectrum of 2MASS 1507$-$1627 (L5).  There is
    a clear minimum in $G$ at 1700 K.}
\end{figure}

As an example of our procedure, Figure \ref{fig:Chi2} shows the $G$
values of all the synthetic spectra in our grid as a function of \teff\
computed for the spectrum of 2MASS 1507$-$1627 (L5).  In this case, the
best fitting synthetic spectrum has the parameters [1700/4.5/2].
However there are additional synthetic spectra that yield similar
minimum $G$ values.  How then can we determine the quality of a given
fit?  Although $G$ is mathematically similar to the $\chi^2$ statistic,
it does not follow a $\chi^2$ distribution\footnote{The $\chi^2$
  distribution with $\nu$ degrees of freedom is defined to be the sum of
  $\nu$ squared standard normal deviates or $\chi^2_{\nu} = \sum z_i^2$,
  where $z_i=(x_i - \mu)/\sigma_i$. Since the models are known to have
  systematic errors, the standardized residuals $z_i = (f_i - C_k
  \mathcal{F}_{k,i})/\sigma_i$ are not standard normal deviates and thus
  $G$ does not follow a $\chi^2$ distribution.}  and as a result, we
cannot employ standard hypothesis testing techniques to assess the
quality of the fits.

We therefore performed a Monte Carlo simulation to determine the range
of synthetic spectra that fit the data given the observational errors.
For each observed spectrum, we generated 2000 simulated data sets
wherein the flux at each wavelength point was randomly drawn from a
Gaussian distribution centered on the observed flux, and with a width
given by the observed variance in the individual pixel.  Since each
0.65$-$14.5 $\mu$m spectrum was constructed by combining together
spectra that were absolutely flux calibrated independently, the error in
any given flux density value also includes a correlated component that
arises from the fact that during the absolute flux calibration process,
all of the flux density values over a range of wavelengths (e.g.,
1$-$2.5 $\mu$m) are scaled by a constant to adjust the overall flux
level to match broad-band photometry.  We therefore scale the simulated
flux density values by 10$^{-0.4 \times K}$ where $K$ is randomly drawn
from a Gaussian distribution centered on zero, and with a width given by
the variance of the photometry (in magnitudes).

A simulated data set s$_i$ is constructed from $m$ spectra, each having
its own $\sigma_{ph}^2$.  For a given spectrum $j$, the simulated
dataset $s_i$ is given by,

\begin{equation}
  s_i = \mathcal{N}(f_i,\sigma_i^2) 10^{-0.4 \times \mathcal{N}(0,\sigma_{ph,j}^2)},
\end{equation}
\noindent
where $\mathcal{N}(\mu,\sigma^2$) represents a value randomly drawn from
a Gaussian distribution with mean $\mu$ and variance $\sigma^2$, and
$\sigma_{ph,j}^2$ is the variance of the photometry used to absolutely
flux calibrate the spectrum $j$.  We then replace $f_i$ with $s_i$ in
Equations 1 and 2 and recompute the best fitting synthetic spectrum.
The quality of the fit between the data and a given synthetic spectrum
is given by \fmc, the fraction of the 2000 simulations in which the
given synthetic spectrum is identified as the best fitting spectrum.  In
the case described above, the synthetic spectrum with the parameters
[1700/4.5/2] has \fmc=1.000 and thus is the best representation of the
data in our grid of synthetic spectra.  Values of \fmc\ much less than
unity imply that more than one synthetic spectrum in the grid fits the
data well.

The best fitting model parameters are given on the model grid points
without interpolation.  Nominally, this would give an internal
uncertainty of half of the grid spacing, or $\pm50\,$K in \teff,
$\pm0.25$ in \logg, and $\pm0.5$ in \fsed.  In practice, the best
fitting models nearly all have \fmc$\sim 1$, which shows that the
internal uncertainties on the fitting parameters are even smaller than
those nominal values.  As we will see in \S\ref{sec:Systematic
  Uncertainties}, our fits of the ultracool dwarf SEDs are affected by
systematic effects that dominate the internal uncertainties and that are
difficult to quantify.  In addition, changes in the models (e.g., the
inclusion of [Fe/H] as an additional parameter) or the use of a
different goodness-of-fit statistic could yield different atmospheric
parameters than we find here.  An accurate determination of physical
parameters of ultracool dwarfs will therefore require the development of
more complete model atmospheres. Nonetheless, trends in the fitted
parameters as revealed by this work should improve our understanding of
the physics of ultracool dwarfs.

\subsection{Fits of the Entire SED} \label{sec:Fits}

\begin{figure} 
\centerline{\hbox{\includegraphics[scale=1.4]{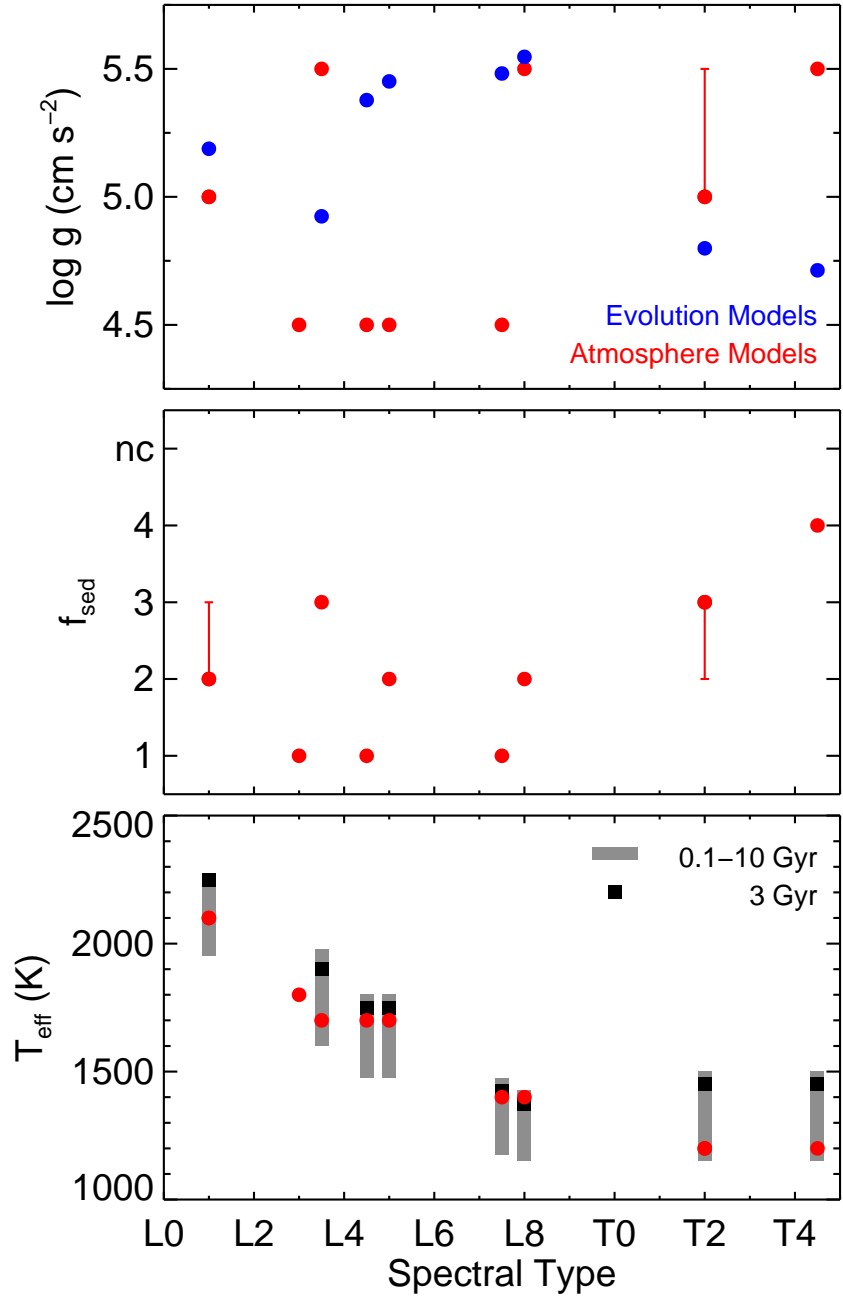}}}
\caption{\label{fig:FullSummary}Summary of the best fitting model
  triplet (\teff/\logg/\fsed) values over the 0.95$-$14.5 $\mu$m
  wavelength range for each dwarf in our sample (\textit{red circles}).
  Errors bars denote the range of parameters for the models with \fmc\
  $>$ 0.1.  In the upper panel, the blue circles are the \logg\ values
  derived using evolutionary models.  In the lower panel, the grey
  regions denote the \teff\ range corresponding to an age range of
  0.1$-$10 Gyr and the black squares denote the \teff\ corresponding to
  3 Gyr \citep{2004AJ....127.3516G}.  All of the \teff\ determinations
  fall within the 0.1$-$10 Gyr \teff\ range.  2MASS 1506$+$1321 (L3)
  lacks a parallax and therefore does not have a \teff\ range plotted.}
\end{figure}

We fit 0.95$-$14.5 $\mu$m spectrum of each dwarf as described in the
previous section.  Although the data extend blueward to 0.65 $\mu$m, the
analysis was limited to wavelengths greater than 0.95 $\mu$m because the
pressure-broadened wing of the resonant \ion{K}{1} doublet (7665, 7699
\AA) that extends to $\sim$0.9 $\mu$m is exceedingly difficult to model
\citep{2003ApJ...583..985B}.  Table \ref{tab:Full} lists the best
fitting model parameters (\teff, \logg, \fsed) for each dwarf in our
sample along with the fraction of the Monte Carlo simulations (\fmc) in
which the given triplet was identified as the best fitting model.  All
models with \fmc\ $>$ 0.1 are listed.  A summary of the derived
parameters is shown in Figure \ref{fig:FullSummary}.  Seven of the nine
dwarfs in our sample have best fitting synthetic spectra with \fmc\
$\sim$1 indicating that they are the only reasonable model spectra in
our grid for these dwarfs.  2MASS 1439$+$1929 (L1) has two model spectra
that differ by 1 in \fsed\ with nearly equal \fmc\ values while SDSS
1254$-$0122 (T2) has two models with \fmc\ $>$ 0.1 that differ by 0.5
dex in \logg\ and 1 in \fsed.

The derived \teff\ values decrease steadily with later spectral type but
the difference in \teff\ between 2MASS 0825$+$2115 (L7.5) and 2MASS
0559$-$1404 (T4.5) is only 200 K which confirms that the transition
between the L and T dwarfs occurs over a narrow range in \teff\ as has
been noted previously \citep[e.g.,
][]{1999ApJ...519..802K,2002ApJ...564..421B,2004AJ....127.3516G}.
\citet{2004AJ....127.3516G} derive the effective temperatures of
$\sim$40 L and T dwarfs using their observed bolometric luminosities and
evolutionary models (see \citet{2002AJ....124.1170D},
\citet{2004AJ....127.2948V}, and \citet{2000ApJ...538..363B} for
alternative \teff\, scales for L and T dwarfs).  Seven of these dwarfs
are included in our sample so we also show in Table \ref{tab:Full} the
\citeauthor{2004AJ....127.3516G} \teff\ values for ages of 0.1, 3, and
10 Gyr.  The agreement between the two sets of values is excellent as
all of our derived values fall within their 0.1 to 10 Gyr ranges.

The comparison of our derived \teff\ values with those of
\citeauthor{2004AJ....127.3516G} implies that if they are single
objects, seven of nine dwarfs in our sample are much younger than 3 Gyr.
Our evolutionary calculations (Marley et al. 2007, in preparation) also
indicate that if we take the derived \teff\ and $g$ values at face
value, five of the L dwarfs would be younger than 0.2 Gyr.  This is
primarily a consequence of the low gravities obtained from fitting the
SEDs.  It seems unlikely that our sample would be biased towards such
young ages given that the mean age of the L and T dwarf field population
is several Gyr \citep{2002AJ....124.1170D,2005ApJ...625..385A}.  This
suggests that fitting the SEDs of ultracool dwarfs is a poor method to
determine gravities.  Fortunately, we can obtain the gravity using an
alternate method.

When the parallax of an object is known, its radius can be determined
using the fitted normalization constant (Eq. \ref{eq:C}) since
$C_k$=($R/d$)$^2$.  With ultracool dwarf evolution sequences,
(\teff,$R$) values can be transformed uniquely into (\teff,$g$) values.
The gravities obtained with this alternate method are given in Table
\ref{tab:Full} and are shown in blue in the upper panel of Figure
\ref{fig:FullSummary}.  The scatter in \logg\ is reduced and nearly all
gravities are above \logg=5.0. Our evolution sequences show that all 9
objects now have ages above 0.3 Gyr. This alternate method only weakly
depends on the relatively subtle variations in spectral morphology
arising from changes in $g$ (Fig. \ref{fig:1800KVariations} and
\ref{fig:1100KVariations}) because it relies primarily on fitting the
absolute flux level of the SED.  The reduced scatter in \logg\ and the
older age distribution of our sample indicate that the method provides
gravity estimates that are more reliable than those obtained by direct
spectral fitting.  Hereafter, we do not compute new model atmospheres
with these new gravities but rather continue to compare the data to
models in our grid with \logg=4.5, 5.0, and 5.5.

We also note that the theoretical evolution of ultracool dwarfs with
cloudy atmospheres is limited to \logg$\le$5.38, although the evolution
with cloudless atmospheres can reach slightly higher gravities of
\logg$\sim$5.48.  Because the mass-\teff-radius relation of brown dwarfs
is well established on theoretical grounds, it provides a firm boundary
to physically plausible combinations of \teff\ and $g$. We find that the
\logg\ values of three of our objects (derived using the alternate
method) exceed the limit for ultracool dwarfs with cloudy atmospheres,
which is an indication that their fitted \teff\ values are overestimated
by $\sim$10$-$20 K.  This result can be understood by noting that the
gravity of brown dwarfs is a very sensitive function of \teff\ for a
fixed \lbol\ (see, for example, Fig. 1 of \citet{2006ApJ...647..552S}).
This is a consequence of the mass-radius relation for degenerate stars
where $R \sim M^{-1/3}$.  Since \lbol\ $\sim$ $R^2$$T_{\mathrm{eff}}^4$,
it follows that at constant \lbol, $g \sim$ $T_{\mathrm{eff}}^{10}$.
From this approximate relation, a 7\% variation in \teff\ changes the
gravity by 0.3 dex.

Two objects in our sample call for a more detailed discussion.  SDSS
1254$-$0122 (T2) and especially 2MASS 0559$-$1404 (T4.5) stand out as
being overluminous for their spectral types \citep{2007ApJ...659..655B}.
It has therefore been suggested that the two objects may be unresolved
binaries
\citep{2004AJ....127.3516G,2004AJ....127.2948V,2006ApJ...647.1393L,2007ApJ...659..655B},
although neither dwarf has been resolved into a binary in
high-resolution imaging \citep{2003ApJ...586..512B,2006ApJS..166..585B}.
We find, as did \citet{2004AJ....127.3516G}, that the two objects share
essentially the same \teff, although we derive \teff=1200 K while
\citeauthor{2004AJ....127.3516G} find \teff=1475 K (assuming ages of 3
Gyr).  If SDSS 1254$-$0122 and 2MASS 0559$-$1404 are equal magnitude
binaries, then the \lbol\ of each component is reduced by a factor of
two and the \citeauthor{2004AJ....127.3516G} \teff\ value becomes
$\sim$1200 K which is in much better agreement with our value.  Given
these new parameters, our evolutionary sequences provide the
corresponding properties for the individual components. To satisfy both
the constraints of component luminosity and the maximum allowed gravity,
\teff\ has to be reduced slightly, but remain within the uncertainties
of our spectral fits.  The properties of the individual components are
given in Table \ref{tab:BinaryTable}.

Whether SDSS 1254$-$0122 and 2MASS 0559$-$1404 are singles or equal mass
binaries, they have essentially the same effective temperature.  The
difference in the appearance of their spectra (T2 versus T4.5), can be
explained as primarily due to a difference in their cloud properties, as
suggested by our fits that indicate the clouds are thinner (larger
\fsed) in 2MASS 0559$-$1404 than in SDSS 1254$-$0122.  Another
possibility would be for SDSS 1254$-$0122 to be a single brown dwarf (or
a strongly unequal-mass binary) and for 2MASS 0559$-$1404 to be an equal
mass binary.  This would make the latter's individual components less
luminous than SDSS 1254$-$0122 (or its primary) and is consistent with
their position in the $M_{\mathrm{bol}}$ vs spectral type diagram
\citep{2007ApJ...659..655B}.

A fair amount of scatter is found in the \fsed\ values of the L dwarfs,
with an indication that \fsed\ increases toward the later T spectral
class, as would be expected if clouds play a decreasing role in shaping
the SED of T dwarfs \citep[e.g.,
][]{2002ApJ...568..335M,2006ApJ...640.1063B}.  Among the L dwarfs in our
sample, \fsed\ ranges from 1 to 3, implying that the cloud opacity
varies appreciably among L dwarfs (bottom panel of Fig. 2) with little,
if any, dependence on the spectral type.  It would be particularly
interesting to establish the distribution of \fsed\ as a function of
spectral type with a larger sample.

\begin{figure} 
\centerline{\hbox{\includegraphics[scale=1.3]{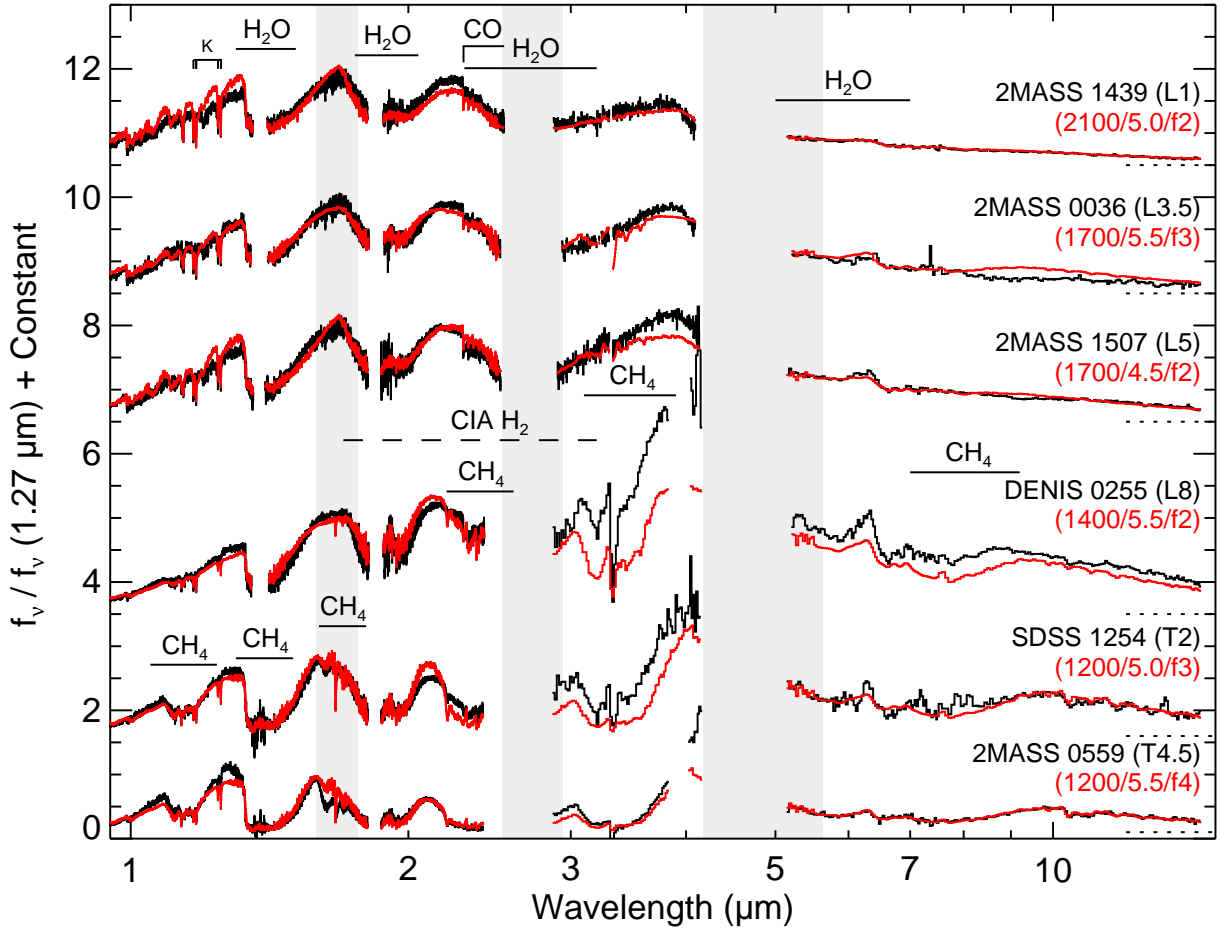}}}
\caption{\label{fig:FullSeq}The 0.95$-$14.5 $\mu$m spectra of 2MASS
  1439$+$1929 (L1), 2MASS 0036$+$1821 (L3.5), 2MASS 1507$-$1627 (L5),
  DENIS 0255$-$4700 (L8), SDSS 1254$-$0122 (T2), and 2MASS 0559$-$1404
  (T4.5) are shown in black.  The data have been normalized to unity at
  1.27 $\mu$m and offset (\textit{dotted lines}) for clarity.  The best
  fitting models (\teff/\logg/\fsed) are shown in red and have been
  normalized and offset with the same constants as the data.  The models
  were multiplied by the constant $C$ (see \S\ref{sec:The Fitting
    Technique}) before normalization to preserve the relative flux
  levels between the data and models.  Regions not included in the fits
  are shown in light grey.}
\end{figure}

Figure \ref{fig:FullSeq} shows the 0.95$-$14.5 $\mu$m spectra of 2MASS
1439$+$1929 (L1), 2MASS 0036$+$1821 (L3.5), 2MASS 1507$-$1627 (L5),
DENIS 0255$-$4700 (L8), SDSS J1254$-$0122 (T2), and 2MASS 0559$-$1404
(T4.5) along with the best fitting model spectra.  The grey regions
indicate wavelength ranges that were excluded from the fits.  The models
fit the data reasonably well at the earliest and latest spectral types,
but do a poorer job of fitting the data in the mid- to late-type L
dwarfs and the early-type T dwarf.  This range of spectral types
corresponds to the L/T transition where condensate clouds have their
largest impact on the SED of ultracool dwarfs.  The relatively poor fits
at the L/T transition most likely reflects the limitations of our simple
cloud model.

In particular, the best fitting model spectra match both the peaks and
depths of the H$_2$O bands centered at 1.4, 1.9, and 2.6 $\mu$m well.
The models do, however, under estimate the $K$-band flux density levels
of 2MASS 1439$+$1929 (L1) and 2MASS 0036$+$1821 (L3.5).  In addition,
the models also over estimate the depth of the $L$-band CH$_4$
absorption bands of both DENIS 0255$-$4700 (L8) and SDSS 1254$-$0122
(T2) as well as the 5.5$-$14.5 $\mu$m flux levels of DENIS 0255$-$4700
(L8).  We defer discussion of the possible causes of these mismatches to
\S\ref{sec:BandbyBandResults}.  Finally, the poor match between the
model spectrum and data of 2MASS 0559$-$1404 in the $H$ band is a result
of the woefully incomplete line list of CH$_4$ at these wavelengths
\citep{2001RvMP...73..719B,freedman07}.

\subsection{Description of Individual  Bands}\label{sec:BandbyBandResults}

\begin{figure} 
\centerline{\hbox{\includegraphics[scale=1.3]{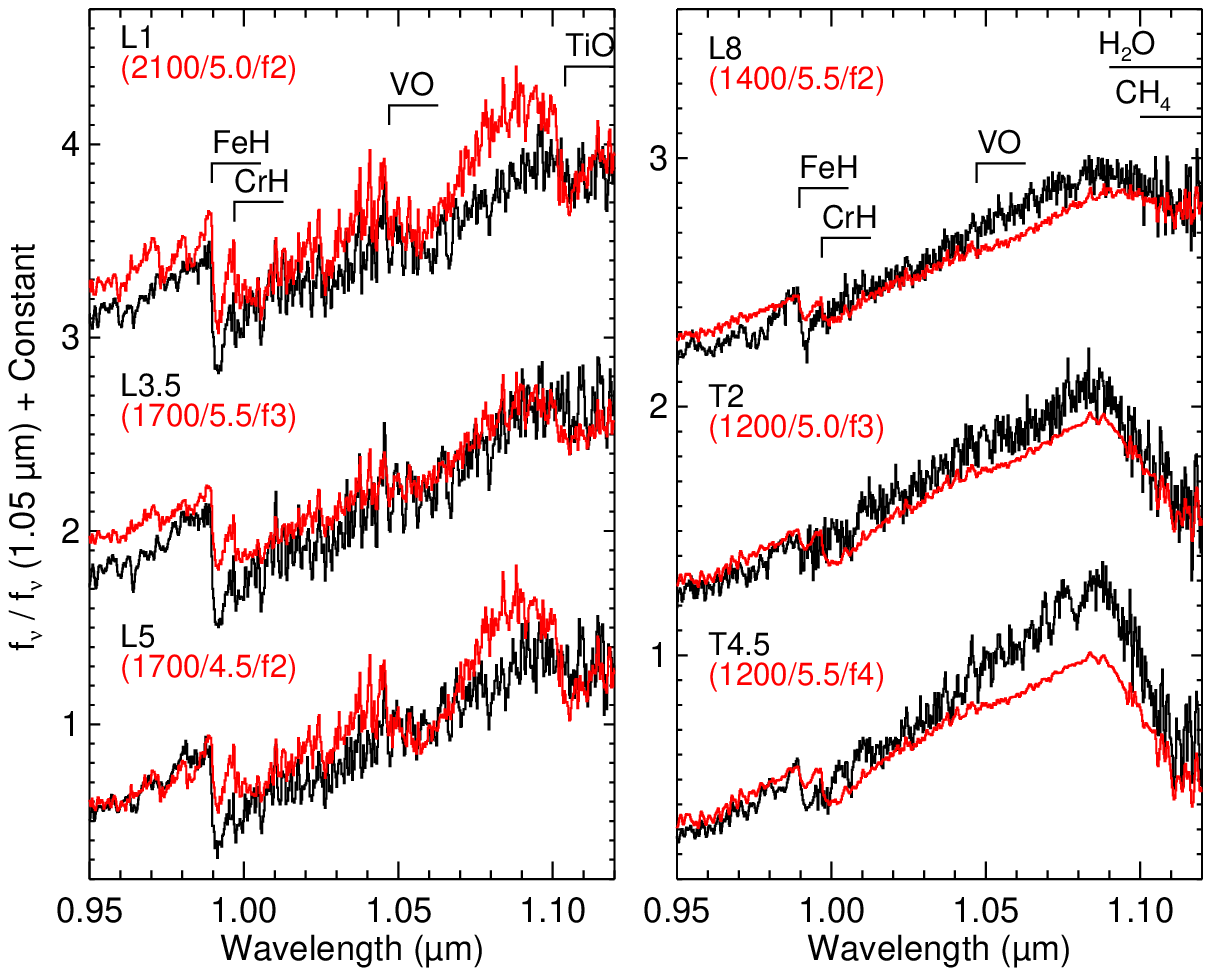}}}
\caption{\label{fig:YSeq1}Same as Figure \ref{fig:FullSeq} except the
  data cover the $Y$ band and were normalized at 1.05 $\mu$m.}
\end{figure}

\begin{figure} 
\centerline{\hbox{\includegraphics[scale=1.3]{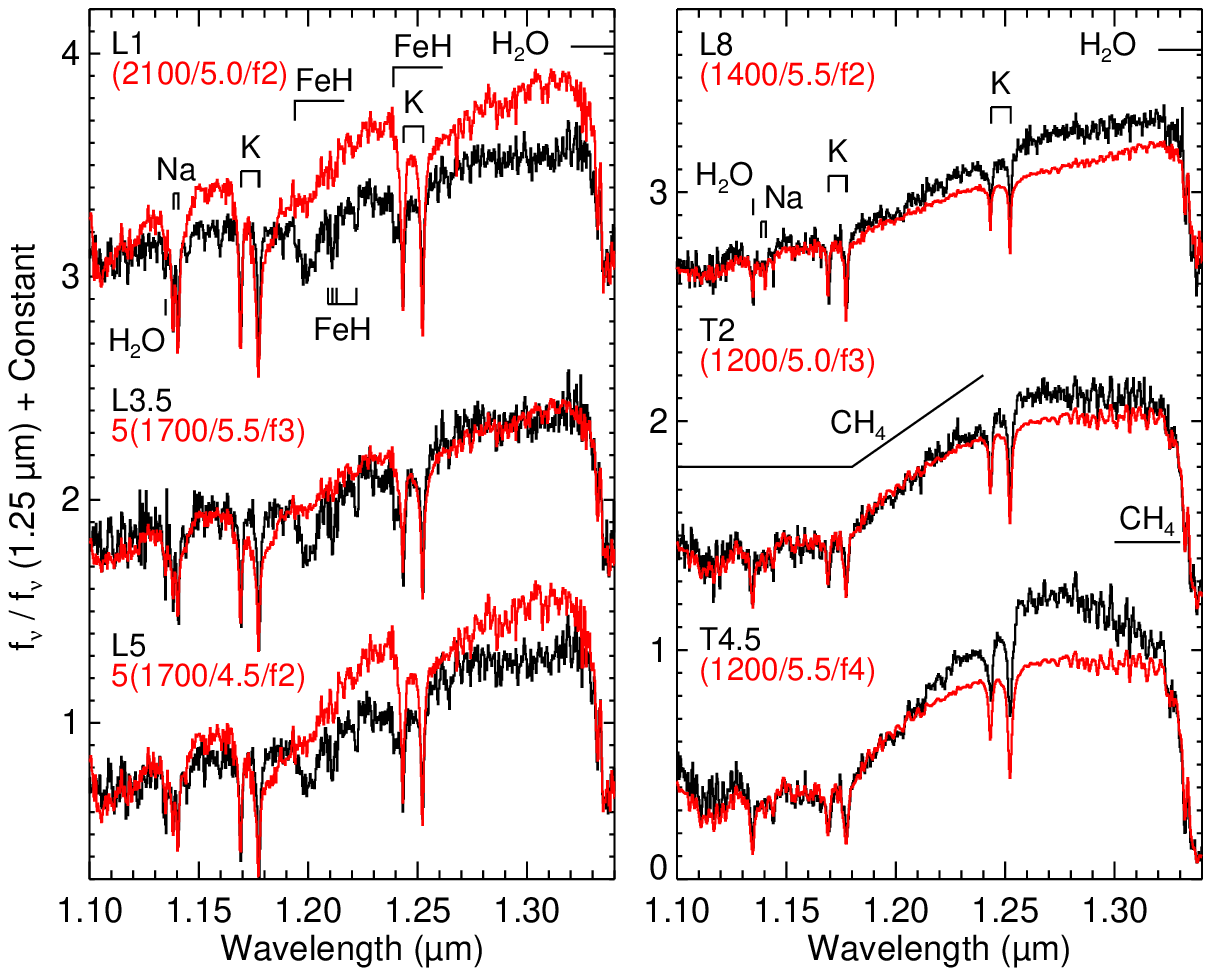}}}
\caption{\label{fig:JSeq1}Same as Figure \ref{fig:FullSeq} except the
  data cover the $J$ band and were normalized at 1.25 $\mu$m.}
\end{figure}

\begin{figure} 
\centerline{\hbox{\includegraphics[scale=1.3]{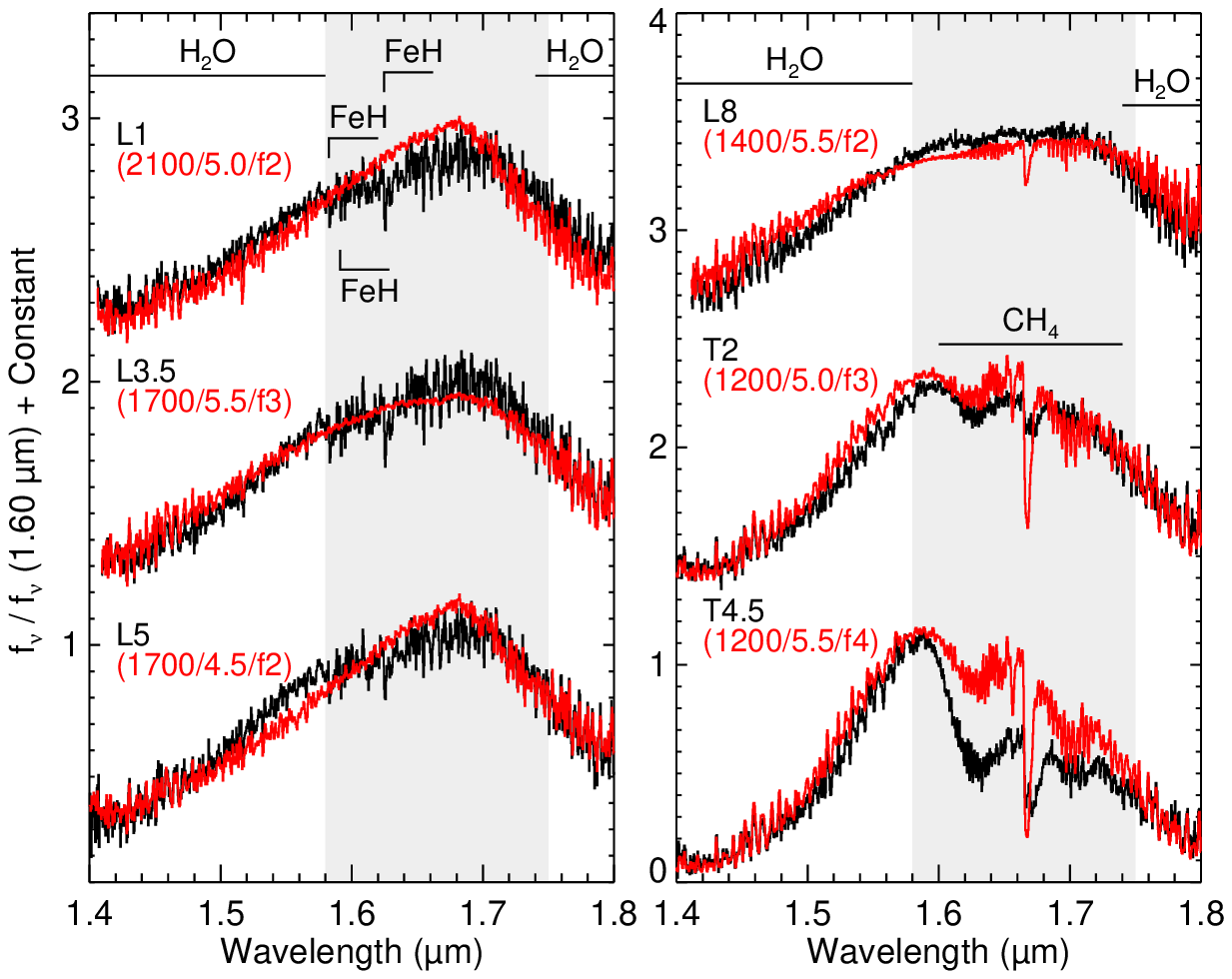}}}
\caption{\label{fig:HSeq1}Same as Figure \ref{fig:FullSeq} except the
  data cover the $H$ band and were normalized at 1.6 $\mu$m.}
\end{figure}

\begin{figure} 
\centerline{\hbox{\includegraphics[scale=1.3]{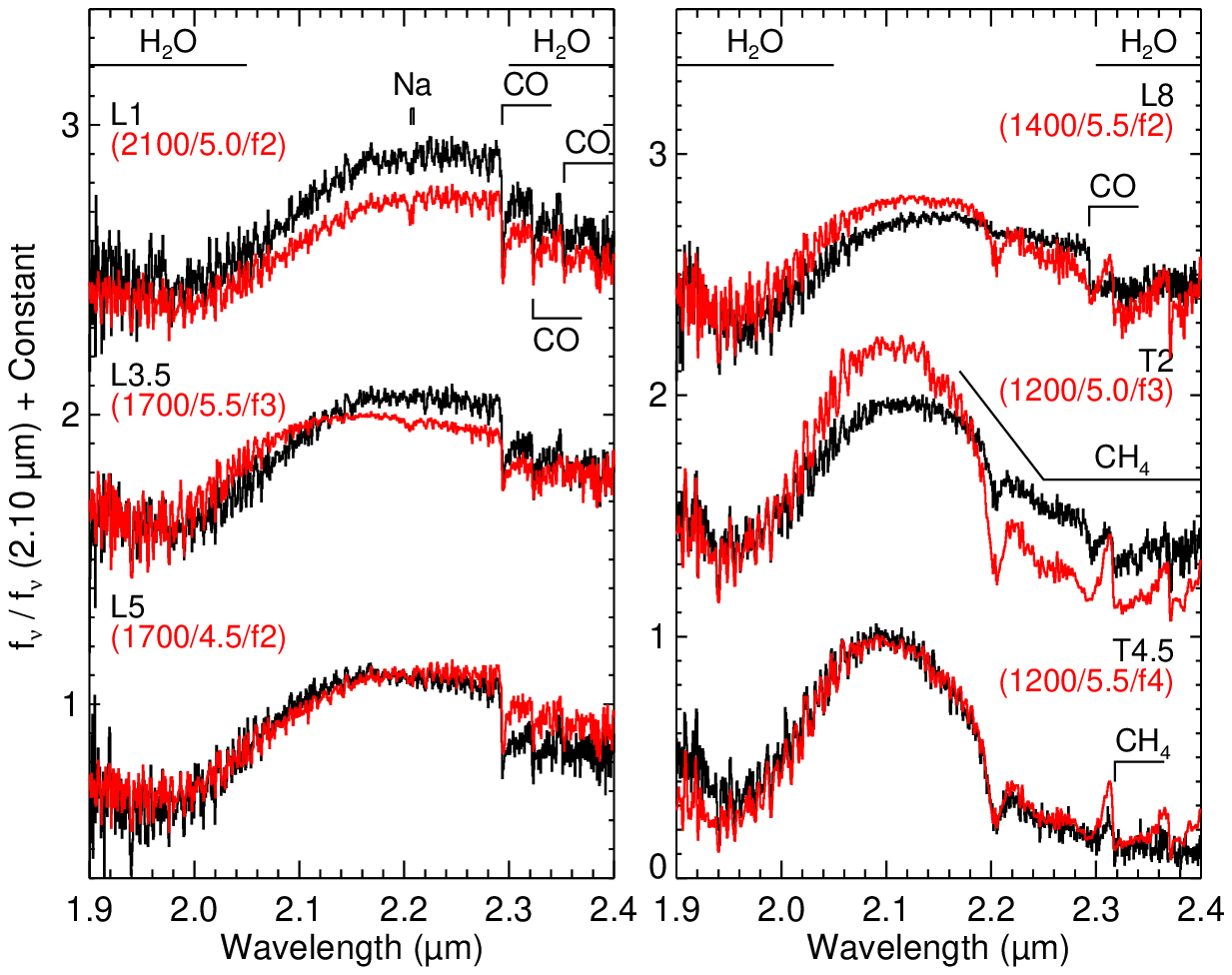}}}
\caption{\label{fig:KSeq1}Same as Figure \ref{fig:FullSeq} except the
  data cover the $K$ band and were normalized at 2.1 $\mu$m.}
\end{figure}

\begin{figure} 
\centerline{\hbox{\includegraphics[scale=1.3]{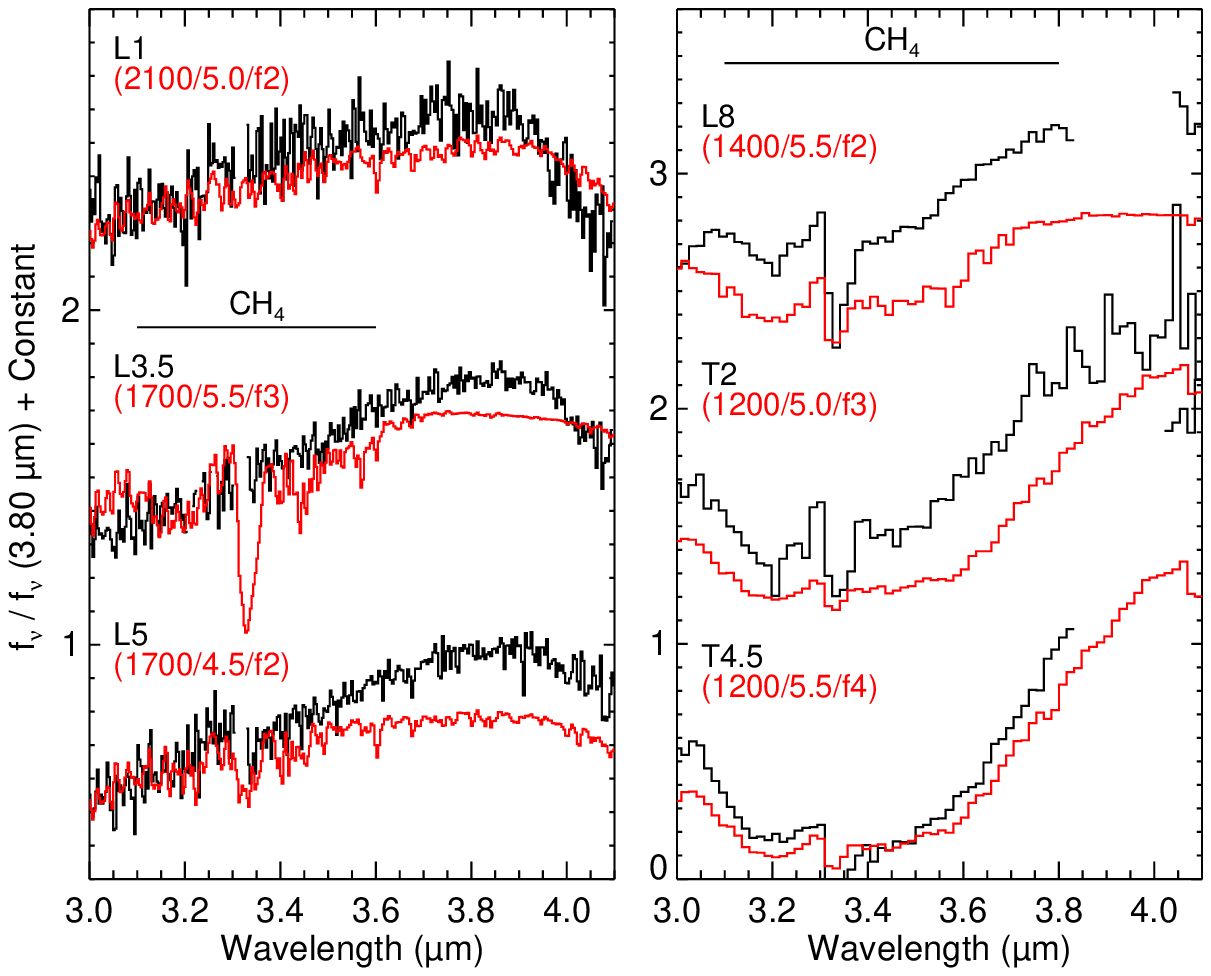}}}
\caption{\label{fig:LSeq1}Same as Figure \ref{fig:FullSeq} except the
  data cover the $L$ band and were normalized at 3.8 $\mu$m.}
\end{figure}

\begin{figure} 
\centerline{\hbox{\includegraphics[scale=1.3]{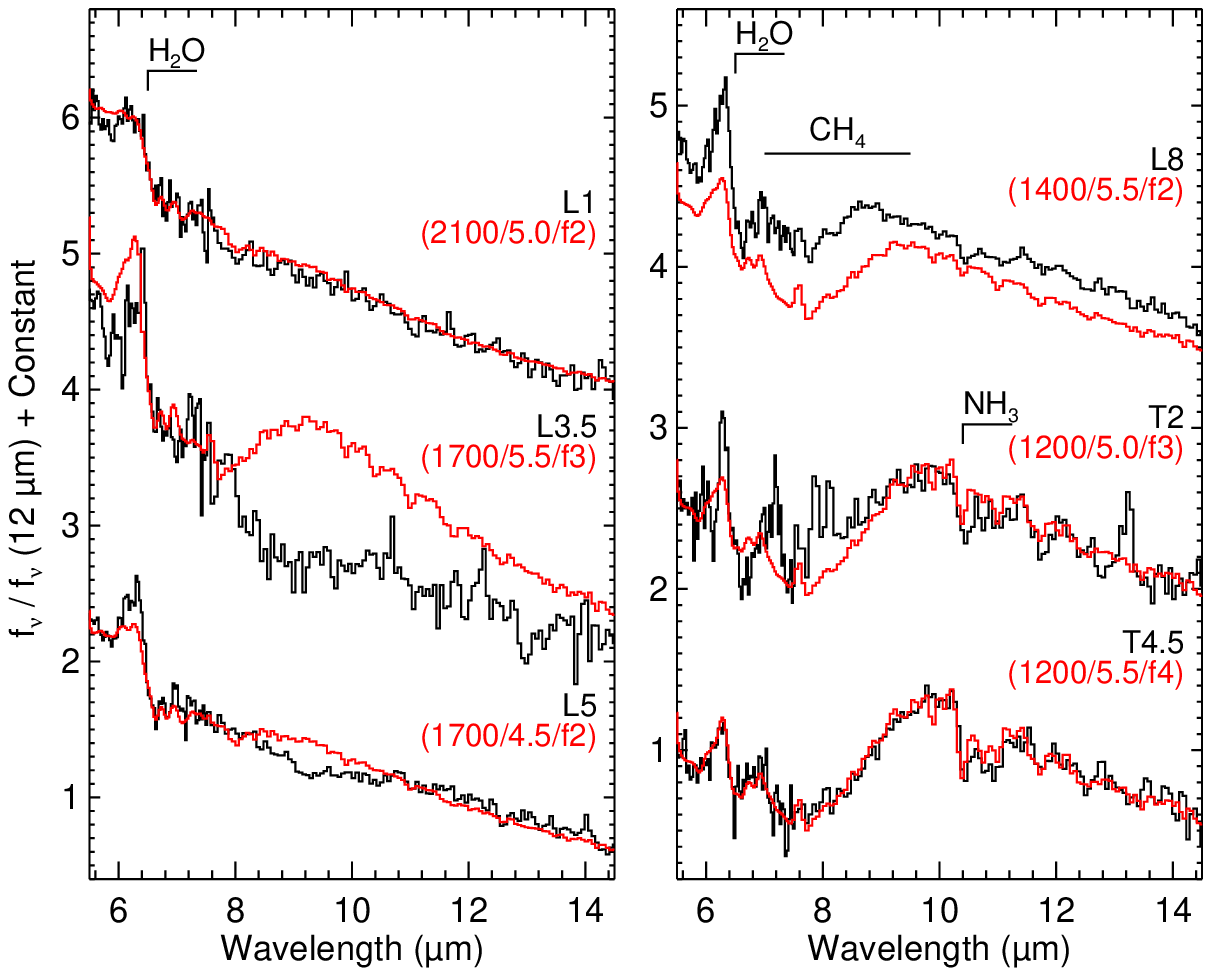}}}
\caption{\label{fig:IRSSeq1}Same as Figure \ref{fig:FullSeq} except the
  data cover the IRS/SL wavelength range and were normalized at 12
  $\mu$m.}
\end{figure}

In the following sections, we describe how well the best fitting
synthetic spectra match the data over narrower wavelength ranges.
Figures \ref{fig:YSeq1} to \ref{fig:IRSSeq1} show the same data and
model spectra presented in Figure \ref{fig:FullSeq} but over the
wavelengths covered by the photometric bandpasses $Y$, $J$, $H$, $K$,
and $L$ and the Short-Low module of the IRS on board \textit{Spitzer}
(hereafter IRS/SL).  Although the most prominent features in the spectra
of the L and T dwarfs are briefly described in each section, more
detailed descriptions can be found elsewhere
\citep{1999ApJ...519..802K,2003ApJ...594..510B,
  2003ApJ...596..561M,2005ApJ...623.1115C,2006ApJ...648..614C}.

\subsubsection{The $Y$ Band}

Figure \ref{fig:YSeq1} shows the spectra and best fitting models over
the wavelength range 0.95 to 1.1 $\mu$m, which roughly corresponds to
the $Y$ band \citep{2002PASP..114..708H}.  The spectra of the L dwarfs
are dominated by absorption features arising from the 0$-$0 band of the
$F$ $^4\Delta$ $-$ $X$ $^4\Delta$ system of FeH.  The strongest FeH
feature is the bandhead at 0.9896 $\mu$m but weak FeH features are seen
in the spectra of the L dwarfs throughout this wavelength range.  The
0$-$0 band of the $A$ $^4\Pi- X$ $^4\Sigma^-$ system of VO centered at
$\sim$1.06 $\mu$m is also present in the spectra of the early-type L
dwarfs and the FeH bandhead at 0.9896 $\mu$m is weakly present in the
spectra of SDSS J1254$-$0122 (T2) and 2MASS 0559$-$1404 (T4.5).
Finally, H$_2$O and CH$_4$ bands at $\lambda >$ 1.09 $\mu$m emerge at
the L/T transition and strengthen through the T sequence.

The synthetic spectra fit the overall shape of the $Y$-band spectra of
the L and T dwarfs reasonably well.  In particular, they match the
strengths of the H$_2$O and CH$_4$ bands in the T dwarf spectra at
$\lambda >$ 1.08 $\mu$m.  However there are a number of mismatches worth
noting.  Firstly, the FeH bandhead in the synthetic spectra is
systematically weaker than in the L dwarf data.  Secondly, all of the
synthetic spectra exhibit the 0$-$1 bandhead of the $A$ $^6\Sigma^+$ $-$
$X$ $^6\Sigma^+$ system of CrH at 0.9969 $\mu$m.  This bandhead is often
identified in the spectra of L dwarfs
\citep[e.g.,][]{1999ApJ...519..802K,2006MNRAS.tmp.1334K} but
\citet{2003ApJ...582.1066C} questioned these identifications given that
many of the absorption features at this wavelength in $R$=2000 spectra
can be associated with FeH.  \citet{2006ApJ...644..497R} recently found
that no CrH absorption features can be identified in high resolution
($R$=31,000) spectra of M and L dwarfs which indicates that CrH is not a
significant absorber at these wavelengths.  Therefore the presence of
this bandhead in the synthetic spectra is inconsistent with the
observations.  The 0$-$1 bandhead of the $\phi$ system of TiO
\citep{1980ApJS...42..241G} at $\sim$1.104 $\mu$m is also too strong in
all of the model spectra.  The T dwarf models exhibit a broad, weak VO
band centered at $\sim$1.06 $\mu$m.  Observationally, this band peaks in
strength at a spectral type of $\sim$L5 and is absent by a spectral type
of L8 \citep{2005ApJ...623.1115C} so the presence of this band in the
models is also inconsistent with the observations.  Finally, the best
fitting model spectrum underestimates the peak flux of the T4.5 dwarf
2MASS 0559$-$1404 because of the presence of the condensate opacity
(\fsed=4).  A cloudless model would match the overall shape of the $Y$
band better (see \S\ref{sec:Y Fit}).

\subsubsection{The $J$ Band}

Figure \ref{fig:JSeq1} shows the spectra and best fitting model spectra
from 1.1 to 1.34 $\mu$m.  This wavelength range contains the most
prominent atomic features in the infrared spectra of L and T dwarfs,
namely the two \ion{K}{1} doublets at $\sim$1.17 $\mu$m and $\sim$1.24
$\mu$m, and the \ion{Na}{1} doublet at 1.14 $\mu$m.  The 0$-$1 and 1$-$2
bands of the $F$ $^4\Delta$ $-$ $X$ $^4\Delta$ system of FeH are also
present in the $J$-band spectra of the L dwarfs.  These two bands
exhibit bandheads at 1.1939 and 1.2389 $\mu$m, respectively, along with
numerous additional weak features from 1.2 to 1.3 $\mu$m.  The 3$\nu_3$
and $\nu_2 + 2\nu_3$ bands of CH$_4$ centered at $\sim$1.15 and
$\sim$1.4 $\mu$m are also seen in the spectra of the T dwarfs.

The model spectra reproduce the depths of the H$_2$O bands at $\lambda
>$1.32 $\mu$m, the onset of the CH$_4$ absorption in the spectra of the
T dwarfs, and the weakening and eventual disappearance of the
\ion{Na}{1} doublet at 1.14 $\mu$m with later spectral type.  However
the model spectra do a poor job of matching the peak flux levels at
$\sim$1.3 $\mu$m of 2MASS 1439$+$1929 (L1), 2MASS 1507$-$1627 (L5), and
2MASS 0559$-$1404 (T4.5).  In addition, the bandheads and other weak FeH
features observed in the $J$-band spectra of the L dwarfs are completely
absent from the models.  This behavior mirrors that in the $Y$-band for
which the FeH band is systematically weaker in the models than in the
data.  The cross-sections of the 0$-$1 and 1$-$2 $\mu$m FeH bands are an
order of magnitude smaller than those of the 0$-$0 FeH band
\citep{2003ApJ...594..651D} so it is not surprising that they are either
very weak or absent in the synthetic spectra.

The models also are a poor match to the early-type L dwarfs near the
1.177 $\mu$m \ion{K}{1} line since the red wing appears much too broad
given the observations.  This apparent line broadening in the model
spectra is actually due to the presence of the 0$-$1 band of the A$-$X
system of VO \citep{1982JMoSp..92..391C}.  Although this band has been
identified in the spectra of young, low-gravity brown dwarfs
\citep{2004ApJ...600.1020M}, it has not been found in the spectra of
field L and T dwarfs and therefore its presence in the model spectrum
implies the opacity of this VO band may be in error.

\subsubsection{The $H$ Band}

Figure \ref{fig:HSeq1} shows the spectra and best fitting synthetic
spectra over the 1.4 to 1.8 $\mu$m wavelength range.  The L dwarf
spectra are shaped primarily by H$_2$O absorption but also exhibit
numerous weak absorption features, including three bandheads, from 1.59
to 1.74 $\mu$m that arise from the $E$ $^4\Pi$ $-$ $A$ $^4\Pi$ system of
FeH.  The 2$\nu_3$ and 2$\nu_2 + \nu_3$ bands of CH$_4$ appear at the
L/T transition and strengthen through the T sequence.  As explained in
\S\ref{sec:The Fitting Technique}, we do not include the 1.58 to 1.75
$\mu$m wavelength range in the fits because the FeH band lacks a line
list and the CH$_4$ line list is incomplete.

The model spectra fit the overall shape of the data reasonably well at
these wavelengths.  In particular, the detailed structure of the H$_2$O
bands at the short and long wavelength ends of the $H$ band are well
matched by the model spectra. The best agreement between the data and
model spectra at these wavelengths are obtained for DENIS 0255$-$4700
(L8) and SDSS 1254$-$0122 (T2) because the FeH band is either weak or
absent from the data and only weak CH$_4$ absorption is present.  This
indicates that the H$_2$O opacity tables are quite good in the $H$ band.

\subsubsection{The $K$ Band} \label{sec:The K Band}

Figure \ref{fig:KSeq1} shows the data and best fitting synthetic spectra
over the 1.9 to 2.4 $\mu$m wavelength range.  The spectra of L and T
dwarfs are shaped primarily by H$_2$O, CO, and CH$_4$ bands at these
wavelengths.  At the earliest spectral types, CO is the dominant carbon
bearing gas and therefore the $\Delta \nu = +$2 CO overtone bands at
$\lambda$ $\gtrsim$ 2.29 $\mu$m are strong.  These CO bands weaken and
disappear with later spectral type as CH$_4$ becomes the dominant
carbon-bearing gas.  Additionally there is a \ion{Na}{1} doublet at 2.26
$\mu$m present in the early-type L dwarfs.

With the exception of 2MASS 1507$-$1627 (L5) and 2MASS 0559$-$1404
(T4.5), the models are a poor fit to the $K$-band spectra.  The models
underestimate the $K$-band flux levels of the 2MASS J1439$+$1929 (L1)
and 2MASS 0036$+$1821 (L3.5) and overestimate the strengths of the
CH$_4$ band in DENIS 0255$-$4700 (L8) and SDSS J1254$-$0122 (T2).  In
addition, a weak CH$_4$ feature can be seen in the synthetic spectrum of
2MASS 0036$+$1821 (L3.5) at $\sim$2.2 $\mu$m that is not observed in
spectra of ultracool dwarfs until a spectral type of $\sim$L8. The
presence of this CH$_4$ feature in the model spectrum of 2MASS
0036$+$1821 is a result of the fact that \fsed=3 for this model. For a
comparison, the best fitting model for 2MASS 1507$-$1627 (L5) has the
same \teff, but \fsed=2, and shows no CH$_4$ feature (the change in $g$
makes little difference in the strength of the CH$_4$ band at this
\teff).  The \fsed=3 model is cooler in the CH$_4$ band formation region
resulting in a higher CH$_4$ abundance and a deeper absorption band.

As described in \S\ref{sec:The Models}, H$_2$ CIA is an important
pressure-sensitive opacity source in the atmospheres of ultracool
dwarfs.  Compared to the absorption of other molecules, (primarily
H$_2$O and CH$_4$), H$_2$ CIA is most significant in the $K$ band. A
lower matallicity decreases the opacity of H$_2$O and CH$_4$ relative to
the H$_2$ CIA continuum, which results in stronger emission in the $Y$,
$J$ and $H$ bands with respect to the $K$ band
\citep{2002ApJ...564..421B}.  In previous work
\citep{2006ApJ...647..552S,2007ApJ...656.1136S}, we found that our solar
metallicity models systematically underestimated the $K$-band fluxes of
late-type T dwarfs while models with [Fe/H] = $+$0.3 fit the $K$-band
peak well.  Therefore the poor match between the data and model spectra
in the $K$ band may indicate that the metallicities of the L and T
dwarfs in our sample differ somewhat form the solar value employed in
our chemical equilibrium calculations.  Another possibility that
explains the mismatch between the spectra DENIS 0255$-$4700 (L8) and
SDSS 1254$-$0122 (T2) and the best fitting models is the cloud
properties since the cloud sedimentation efficiency \fsed\ changes from
2 to 3 to 4 from L8 to T4.5.

\subsubsection{The $L$ Band}\label{sec:The L Band}

Figure \ref{fig:LSeq1} shows the spectra and best fitting synthetic
spectra over the 3.0 to 4.1 $\mu$m wavelength range.  The spectra are
dominated by weak H$_2$O features and the $\nu_3$ fundamental band of
CH$_4$.  The Q-branch ($\sim$3.3 $\mu$m) is the first CH$_4$ feature to
emerge in the spectra of the L dwarfs and does so at a spectral type of
$\sim$L5.  With increasing spectral type, the P and the R branches
emerge and strengthen through the spectral sequence until the band is
saturated in the late-type T dwarfs.

The model spectra do not match the data particularly well in the $L$
band.  The most egregious mismatch is 2MASS 0036$+$1821 (L3.5) for which
the model shows a very strong CH$_4$ band that is absent in the data.
The presence of the CH$_4$ band is due to the fact that the best fitting
model has \fsed=3, as in the $K$ band (see \S\ref{sec:The L Band}).  The
average flux density levels of the models also appear low for 2MASS
1507$-$1627 (L5), DENIS 0255$-$4700 (L8), SDSS J1254$-$0122 (T2), and
2MASS 0559$-$1404 (T4.5).  In addition, the model for 2MASS J0559$-$1404
does not match the peak of the observed spectrum at $\sim$4.1 $\mu$m.
Generally, in objects where the CH$_4$ absorption is too strong in the
$L$ band, a corresponding mismatch occurs in the $K$ and IRS/SL bands
(L8 and T2).  The line lists for both H$_2$O and CH$_4$ are reasonably
complete at these wavelengths so the most likely explanation for the
mismatch is that the region of the model atmosphere from which the $K$
and $L$ band and IRS/SL flux emerges is too cool, possibly as a result
of the condensate cloud model.


\subsubsection{IRS/SL}

Figure \ref{fig:IRSSeq1} shows the data and best fitting models over the
5.5 to 14.5 $\mu$m wavelength range covered by the Short-Low module of
the IRS \citep{2004ApJS..154...18H}.  H$_2$O absorption features arising
from the $\nu_2$ fundamental band and 2$\nu_2 - \nu_2$ overtone band are
present throughout this wavelength range; the break in the spectra at
6.5 $\mu$m is an H$_2$O feature.  The $\nu_4$ fundamental band of CH$_4$
centered at $\sim$7.65 $\mu$m emerges in the spectra of late-type L
dwarfs and the $\nu_2$ fundamental band of NH$_3$ centered at $\sim$10.5
$\mu$m appears in the spectra of the T dwarfs.

With the exception of 2MASS 0036$+$1821 (L3.5), the synthetic spectra
match the data reasonably well.  In particular, the strengths of the
H$_2$O band at 6.5 $\mu$m, and the CH$_4$ band at 7.8 $\mu$m is well
matched by the model spectra. The flux level of the synthetic spectrum
of DENIS 0255$-$4700 (L8) is low however, and there is a slight mismatch
between the model spectrum and 2MASS 1507$-$1627 (L5) at $\sim$9 $\mu$m
that we discuss further in \S\ref{sec:IRS/SL Fit}.  The mismatch between
the model and the spectrum of 2MASS 0036$+$1821 is due to the fact that
the model has \fsed=3 (see \S\ref{sec:The K Band} and \S\ref{sec:The L
  Band}).

\subsection{Fits to Individual Photometric  Bands}\label{sec:BandbyBandFits}

The L and T dwarf spectra studied herein cover a broad range of
wavelengths while most published spectra cover only the red optical and
near-infrared wavelengths \citep[e.g.,
][]{2003AJ....126.2421C,2004AJ....127.3553K} or even a single
photometric band
\citep[e.g.,][]{2005A&A...435L..13N,2006ApJ...651.1166M}.  It is
therefore appropriate to investigate how well the atmospheric parameters
of L and T dwarfs can be determined by fitting L and T dwarf spectra
over narrower wavelengths ranges.  In addition, we can also estimate the
systematic errors in both the models and spectral fitting procedure by
comparing how well the models fit the spectral features over narrower
wavelength ranges as compared to the global 0.95$-$14.5 $\mu$m fits.

We therefore fit the data over the wavelength ranges described in
\S\ref{sec:BandbyBandResults} as well as over the 0.95$-$2.5 $\mu$m
wavelength range.  For the IRS/SL and $Y$-, $J$-, $K$-, and $L$-band
fits, we set $w_i$=1 for all $i$.  Table \ref{tab:MicroFits} lists the
results of the fits and Figure \ref{fig:TeffSummary}, which is similar
to the lower panel of Figure \ref{fig:FullSummary}, shows the
corresponding \teff\ values for each dwarf.  Figures
\ref{fig:YSeq2}$-$\ref{fig:KSeq2} and
\ref{fig:LSeq2}$-$\ref{fig:YJHKSeq2} show the spectra and best fitting
models (\textit{blue}) along with the best fitting models obtained by
fitting the 0.95$-$14.5 $\mu$m (\textit{red}, same as Fig.
\ref{fig:FullSeq}).

As expected, the model spectra fit the data much better over these
narrower wavelength ranges.  In most cases, remarkably good fits are
obtained.  However, the \teff\ values derived from narrower wavelength
ranges do not agree with those derived by fitting the 0.95$-$14.5 $\mu$m
SEDs.  Particularly poor \teff\ values are derived by fitting the $L$
band and IRS/SL spectra of the early- to mid-type L dwarfs (see Figure
\ref{fig:TeffSummary}).  Below, we describe how the model spectra fit
the data over the narrower wavelength ranges in more detail.  Since the
majority of the $H$ band is not included in the fitting process, (see
\S\ref{sec:The Fitting Technique}), we do not present a detailed
description of that wavelength region below.

\subsubsection{$Y$ Band Fits} \label{sec:Y Fit}

Figure \ref{fig:YSeq2} shows the spectra and best fitting model spectra
over the $Y$ band.  The fits reproduce the overall shape of the $Y$ band
data considerably better than 0.95$-$14.5 $\mu$m fits for all spectral
types.  The remaining descrepencies highlight problems with specific
molecular bands.  The model FeH bandheads at 0.9896 $\mu$m are still too
weak, the 0.8611 $\mu$m CrH bandheads are still present, and the TiO
bandheads at $\sim$1.104 $\mu$m are still too strong.  Indeed the CrH
bandhead is present in all the model spectra shown in Figure
\ref{fig:YSeq2}.  The strength of the $\sim$1.06 $\mu$m VO band is,
however, much better matched when fitting the $Y$ band data alone.  The
peak flux of 2MASS 0559$-$1404 (T4.5) is also better matched when
fitting the $Y$ band data alone but the synthetic spectrum does exhibit
the VO band that is absent from the data.

As shown in Figure \ref{fig:TeffSummary}, six of the nine dwarfs have
derived \teff\ values that fall within the \citet{2004AJ....127.3516G}
effective temperature range.  However the derived \teff\ values for the
early- to mid-type L dwarfs (L1 to L5) differ by no more than 200 K
while the \citet{2004AJ....127.3516G} 3-Gyr effective temperatures and
the \teff\ values derived by fitting the 0.95$-$14.5 $\mu$m spectra vary
by 500 K over the same spectral type range which indicates the $Y$ band
is not a sensitive indicator of effective temperature at these spectral
types.

\subsubsection{$J$ Band Fits} \label{sec:J Fit}

Figure \ref{fig:JSeq2} shows the spectra and best fitting models over
the 1.1 to 1.34 $\mu$m wavelength range.  Fits to the $J$-band spectra
alone significantly improve for those dwarfs where the global fit was
rather poor, i.e., for 2MASS 1439$+$1929 (L1), 2MASS 1507$-$1627 (L5),
and 2MASS 0559$-$1404 (T4.5).  The FeH features are still absent from
the models in the early- to mid-type L dwarfs, but overall, very good
fits in the $J$ band can be obtained for the whole spectral sequence.
Except for 2MASS 1439$+$1929 (L1) and 2MASS 0036$+$1821 (L3.5), the
\teff\ values obtained with the $J$ band are consistent with those
derived from fitting the full SED.

\subsubsection{$K$ Band Fits} \label{sec:K Fit}

Figure \ref{fig:KSeq2} shows the spectra and best fitting models over
the 1.9 to 2.4 $\mu$m wavelength range.  The improvement in the quality
of the fits is substantial when the $K$-band spectra are fitted alone.
Indeed the agreement between the synthetic spectra and data is at its
highest at these wavelengths; the detailed structure of the H$_2$O, CO,
and CH$_4$ absorption features throughout this wavelength range are
almost perfectly reproduced by the models (see Figure \ref{fig:KZoom}).
However the improvement in the fits comes from a change in \teff\ of up
to 300 K and the $K$-band \teff\ values show more scatter.
\citet{2007ApJ...666.1198B} have fitted high resolution ($R$ $\approx$
50000) 2.296$-$2.308 $\mu$m spectra of nine L dwarfs spanning spectral
types L0 to L6 using our model atmospheres and also find good agreement
between the data and models.

\subsubsection{$L$ Band Fits} \label{sec:L Fit}

Figure \ref{fig:LSeq2} shows the spectra and best fitting models over
the 3.0 to 4.1 $\mu$m wavelength range.  There is a marked improvement
in the quality of the fits when the $L$-band spectra are fitted
separately.  The early- to mid-type L dwarfs no longer show the deep
Q-branch CH$_4$ absorption and the CH$_4$ band in DENIS 0255$-$4700
(L8), SDSS J1254$-$0122 (T2), and 2MASS 0559$-$1404 (T4.5) is well
matched by the synthetic spectra.  The model spectrum now fits the peak
flux in the spectrum of 2MASS 0559$-$1404 at 4.1 $\mu$m well.  The
derived \teff\ values are systematically hotter than those we obtained
from the other bands by up to 500 K for all the dwarfs except 2MASS
0559$-$1404 (Fig.  \ref{fig:TeffSummary}) which indicates that the $L$
band is a poor spectral region from which to derive atmospheric
parameters of L and T dwarfs.  The low value of \fmc\ for some of the
dwarfs in our sample (Table \ref{tab:MicroFits}) reinforces this
conclusion.

\subsubsection{IRS/SL Fits} \label{sec:IRS/SL Fit}

Figure \ref{fig:IRSSeq2} shows the data and best fitting models over the
5.5 to 14.5 $\mu$m wavelength range.  There is also a marked improvement
in the fits over this wavelength range since the strengths of the
H$_2$O, CH$_4$, and NH$_3$ bands are well reproduced by the model
spectra.  As first noted by \citet{2006ApJ...648..614C}, there is a
mismatch between the data and the models from 9 to 11 $\mu$m in the
spectra of the mid-type L dwarfs.  This mismatch is most prominent in
the spectrum of 2MASS 2224$-$0148 \citep[c.f., Fig. 9,
][]{2006ApJ...648..614C} but can also been seen weakly in 2MASS
0036$+$1821 (L3.5) and 2MASS 1507$-$1627 (L5).
\citeauthor{2006ApJ...648..614C} have tentatively identified this
feature as arising from the Si-O stretching mode of small silicate
grains that give rise to the classic 10 $\mu$m silicate feature.
\citet{2006A&A...451L...9H} suggest that the absorption feature could
arise from quartz (SiO$_2$) grains, although this is not expected from
equilibrium chemistry.

The derived \teff\ values for the early- to mid-type L dwarfs (Fig.
\ref{fig:TeffSummary}) are well above the globally determined values
(Fig. \ref{fig:FullSummary}) by up to 700 K.  For four dwarfs, the best
fit is found in the high-\teff\ high-$g$ corner of our model grid
(\teff=2400 K, \logg=5.5) and should be considered with caution.  The
IRS/SL spectral range gives \teff\ values that behave much like those
obtained by fitting the $L$ band.  Both the $L$ and IRS/SL bands probe
regions of the atmosphere with the the lowest pressures and
temperatures.  The tendency of the $L$ and IRS/SL band fits to produce
high \teff\ values combined with the fact that the global 0.9$-$14.5
$\mu$m fits tend to over predict the depth of the 3.3 and 7.8 $\mu$m
methane bands implies that the tops of the model atmospheres may be too
cold.  This could be simply due to flux redistribution caused by
inadequate CH$_4$ opacities at $\lambda <$ 1.7 $\mu$m or a different
vertical distribution of opacity within the cloud.  On the other hand,
atmosphere models also tend to under-predict the temperatures of the
middle and upper atmospheres of solar system giant planets
\citep[e.g.,][]{1999Icar..138..268M}, even when accounting for the
effects of incident radiation and photochemistry. One possible
explanation for this discrepancy is that upwardly propagating waves
launched within the convection region break and deposit their energy in
the lower pressure regions of the atmosphere (e.g., the discussion in
\citet{2005Icar..173..185Y} for Jupiter).  An exploration of the
possible effects of such waves in ultracool dwarf atmospheres would be
of interest.

\citet{2006ApJ...647..552S,2007ApJ...656.1136S} have found that the
IRS/SL spectra of Gl 570D (T7.5) and 2MASS J04151954$-$0935066 (T7) can
be modeled adequately only by reducing the abundance of NH$_3$ by about
an order magnitude from that predicted by chemical equilibrium.  This is
attributed the vertical motions of the gas within the atmospheres which
prevents the abundances of N$_2$ and NH$_3$ from reaching their expected
equilibrium values \citep{2002Icar..155..393L}.  The models presented
herein do not include the effects of non-equilibrium chemistry.  We note
that an equilibrium model fits the IRS/SL spectrum of 2MASS 0559$-$1404
(T4.5) well.  This is a consequence of its higher \teff, which decreases
the effect of vertical transport on the chemistry of NH$_3$.
Nevertheless, a non-equilibrium model provides a better fit to the
NH$_3$ features of the IRS Short High spectrum ($R$$\approx$600) of this
object at a 2-$\sigma$ level \citep{2007ApJ...662.1245M}.

\subsubsection{0.95$-$2.5 $\mu$m Fits} \label{sec:Near Infrared Fit}

Figure \ref{fig:YJHKSeq2} shows the data and best fitting models over
the 0.95 to 2.5 $\mu$m wavelength range.  With the exceptions of 2MASS
1439$+$1929 (L1) and SDSS 1254$-$0122 (T2), fitting the near-infrared
wavelengths alone does not dramatically improve the quality of the fits.
Indeed the same model ([1700/5.5/3]) fits both the near-infrared and
0.95$-$14.5 $\mu$m spectra of 2MASS 0036$+$1821 (L3.5).  Still, the
depths of the H$_2$O and CH$_4$ bands in the remaining dwarfs are better
matched when fitting the near-infrared spectra alone.  Note also that
the near-infrared spectra of 2MASS 0036$+$1821 (L3.5) and 2MASS
1507$-$1627 (L5) are fit by the same model (1700/5.5/f3).  The derived
\teff\ values do not match the 0.95$-$14.5 $\mu$m \teff\ values well,
particularly for the early- to mid-type L dwarfs.  This implies that
spectra at $\lambda \gtrsim$ 2.5 $\mu$m are necessary in order to derive
\teff\ values for early- to mid-type L dwarfs.  This is unfortunate
since most published spectra of L dwarfs rarely extend beyond 2.5
$\mu$m.

\subsection{The L/T Transition}

The transition from the L to T dwarfs is marked by a rapid change in
near-infrared color from the very red late-type L dwarfs
($J-K$$\sim$2.5) to the very blue T dwarfs ($J-K$$\sim$$-$1) and an
increase in the emergent flux in the 1 $\mu$m region
\citep{2002AJ....124.1170D,2003AJ....126..975T,2004AJ....127.2948V}.
The latter trend was unexpected because theoretical \lbol\ declines as a
brown dwarf cools.  The 0.9896 $\mu$m FeH bandhead also strengthens in
the spectra of the early- to mid-type T dwarfs after weakening
considerably in the spectra of the late-type L dwarfs
\citep{2002ApJ...571L.151B,2003ApJ...596..561M}.  Furthermore, the
transition occurs over a very small \teff\ range, perhaps only 100 to
200 K \citep{2004AJ....127.3516G}.  All current atmospheric and
evolutionary models
\citep{2002ApJ...568..335M,2005ApJ...621.1033T,2006ApJ...640.1063B} fail
to reproduce these observations.

A number of possible solutions have been proposed including the break up
of the condensate cloud decks
\citep{2001ApJ...556..872A,2002ApJ...571L.151B}, a rapid change in the
efficiency of condensate sedimentation \citep{2004AJ....127.3553K}, or
even ``crypto-binarity'' \citep{2006ApJ...640.1063B}.  Unresolved
binaries do play a role in exaggerating the magnitude of the 1 $\mu$m
brightening in the field population
\citep{2006ApJS..166..585B,2007ApJ...659..655B,2006ApJ...647.1393L} but
the brightening appears to be intrinsic to the evolution of brown dwarf
atmospheres since it has also been observed in binary systems composed
of early-type and mid-type T dwarfs.  In these systems, whose components
presumably have the same metallicity and are coeval, the component with
the \textit{later} spectral type is \textit{brighter} in the $J$ band
\citep{2005ApJ...634L.177B,2006ApJ...647.1393L}.

Although the models involving changes in the condensate cloud properties
are both consistent with the color magnitude diagrams of field L and T
dwarfs at the L/T transition
\citep{2002ApJ...571L.151B,2004AJ....127.3553K}, the model spectra have
yet to be directly compared to spectroscopic observations.  Indeed an
error in the TiO line list used in a previous generation of our models
(see \S\ref{sec:The Models}) produced emergent spectra with TiO bands at
near-infrared wavelengths that were much too strong, but whose $J$- and
$K$-band magnitudes nevertheless matched the colors of L dwarfs
relatively well.  Unfortunately, self-consistent models that include the
break-up of the condensate clouds decks are not yet available so we can
only compare models with changes in the sedimentation efficiency \fsed.

Four of the dwarfs in our sample span the L/T transition: 2MASS
0825$+$2115 (L7.5), DENIS 0255$-$4700 (L8), SDSS J1254$-$0122 (T2), and
2MASS 0559$-$1404 (T4.5).  As can be seen in Figure \ref{fig:FullSeq}
and Table \ref{tab:Full}, the effective temperatures derived from
fitting the 0.9$-$14.5 $\mu$m spectra of these four dwarfs decrease
slightly from 1400 K to 1200 K and the clouds become gradually thinner
(\fsed\ increases) toward later spectral types. Our fits of the SEDs of
late-type L and early-type T dwarfs therefore support the concept of a
rapid decrease in the overall cloud opacity over a narrow range in
\teff.

\citet{2002ApJ...571L.151B} argue that the resurgence of the 0.9896
$\mu$m FeH bandhead in the spectra of early- to mid-type T dwarfs
\citep[see also][]{2003ApJ...596..561M} is consistent with the cloud
breakup hypothesis.  In the absence of such a breakup, radiation at
$\sim$1 $\mu$m only emerges from atmospheric layers above the Fe cloud
deck in which gaseous FeH has almost completely been depleted into solid
or liquid Fe.  If holes are present in the cloud decks, radiation from
atmospheric layers below the Fe cloud deck where FeH has not been
depleted can escape and as a result, the FeH bandhead is observed in the
spectra.

Figure \ref{fig:FeH} shows a sequence of models with \teff=1400 K,
\logg=5.0, and \fsed=1, 2, 3, 4, and nc.  The strength of the FeH
bandhead also increases in strength with increasing sedimentation
efficiency.  For a fixed $P$/$T$ profile, the FeH column density above
the Fe cloud base ($T$ $\sim$ 2000 K) is set by the vapor pressure of Fe
and thus is independent of both metallicity and \fsed\footnote{The
  formation of FeH gas can be written as Fe (gas) + 0.5 H$_2$ = FeH
  (gas).  If Fe is condensed, the partial pressure of Fe (gas) is fixed
  by the vapor pressure of the iron and is independent of the total
  amount of condensed phase.  Since the partial pressure of H$_2$ is
  essentially a constant (at a given total pressure), the partial
  pressure of FeH is also fixed. Therefore once Fe condenses, the
  partial pressure, and thus column density, of FeH above the Fe cloud
  base is independent of both metallicity and \fsed for a fixed $P$/$T$
  profile.}.  In the \fsed=1 model, the $\tau$ $\sim$ 2/3 level is high
in the cloud ($T$ $\approx$ 1500 K) but as \fsed\ increases, the Fe
cloud becomes thinner and the $\tau$ $\sim$ 2/3 level is reached deeper
in the atmosphere.  As a result, the column density of FeH that is
observed increases resulting in an increase in the band strength of FeH.
Therefore a change in sedimentation efficiency at a fixed \teff\ can
also explain the resurgence of the FeH bandhead across the L/T
transition.

\subsection{Very Red L Dwarfs} \label{sec:Very Red L Dwarfs}

As can be seen in Figure \ref{fig:colors}, there is significant scatter
($\sim$0.5$-$1 mag) in the near-infrared colors of L and T dwarfs at a
given spectral type.  It has been suggested that this spread arises from
variations in the condensate cloud properties
\citep{2004AJ....127.3553K} or surface gravity
\citep{2006ApJ...640.1063B}, although unresolved binaries also
contribute to the scatter \citep{2005ApJ...634..616L}.  In our models,
both surface gravity and cloud sedimentation efficiency affect the
near-infrared colors, but the latter dominates for the mid- to late-type
L dwarfs.  Two pairs of objects in our sample, 2MASS 2224$-$1058/2MASS
1507$-$1627 and 2MASS J0825$+$2115/DENIS J0255$-$4700, have identical
spectral types within the errors but markedly different $J-K_s$ colors:
2.05/1.52 and 2.07/1.69, respectively.  As a result, they are ideal for
studying the dependence of surface gravity and the condensate cloud
properties on the colors of L dwarfs at a given spectral type.

The best fitting model spectra over the 0.95 to 14.5 $\mu$m wavelength
range for 2MASS 2224$-$0158 and 2MASS 1507$-$1627 have [1700/4.5/1] and
[1700/4.5/2], respectively.  The two objects have the same \logg\ and
\teff\ values, but the redder object, 2MASS 2224$-$0158 has an \fsed\
value of 1 while the bluer 2MASS J1507$-$1627 has \fsed=2.  Condensate
clouds with \fsed=1 are thicker and have smaller particles and therefore
contribute more opacity to the atmosphere (which result in redder
$J-K_S$ colors (see \S\ref{sec:The Models})).  \citet{burgasser07a} have
also noted that the broad absorption feature seen in the IRS spectrum of
2MASS J2224$-$0158, which \citet{2006ApJ...648..614C} have tentatively
identified as arising from small silicate grains in the condensate
clouds (see \S\ref{sec:IRS/SL Fit}), is much weaker in the IRS spectrum
of the bluer 2MASS 1507$-$1627.  This is consistent with the condensate
clouds being responsible for the very red color of 2MASS 2224$-$0158.

Similarly, the best fitting model spectra over the 0.95 to 14.5 $\mu$m
wavelength range for 2MASS 0825$+$2115 and DENIS 0255$-$4700 have
[1400/4.5/1] and [1400/5.5/2], respectively.  These two objects have
identical \teff\ and different \logg\ values but as above, the redder
object, 2MASS 0825$+$2115, has \fsed=1.  These latter two objects show
how both surface gravity and the condensate clouds play a role in
controlling the colors of L dwarfs.

\section{Systematic Uncertainties} \label{sec:Systematic Uncertainties}

Both the $G$ statistic and the Monte Carlo simulations used in this
analysis explicitly account for the random errors in the data.  However
it is important to also understand any systematic errors that can affect
the analysis.  Below, we discuss two such systematic errors, the
absolute flux calibration process and the choice of the weights in the
$G$ statistic.

\subsection{Absolute Flux Calibration} \label{sec:Absolute Flux Calibration}

The near-infrared spectra used in this work were obtained with SpeX on
the NASA IRTF.  The 0.8$-$2.5 $\mu$m wavelength range is covered in six
cross-dispersed orders, and as a result the spectra from the various
orders are merged into a single spectrum by scaling the spectra to
ensure that the flux density levels match in regions where the spectra
overlap in wavelength.  Unfortunately there is no wavelength overlap
between the $H$ and $K$ bands and therefore no scaling could be
performed.  The merged 0.8$-$2.5 $\mu$m spectra were then absolutely
flux calibrated by deriving a \textit{single} scale factor based on the
2MASS $JHK_s$ photometry.  \citet{2005ApJ...623.1115C} showed that, on
average, the synthetic colors ($J-H$, $H-K_s$, and $J-K_s$) of their
entire sample of M, L, and T dwarfs match the observed 2MASS colors
within the errors.  However in some cases, we find deviations between
the synthetic and observed colors of up to 2 $\sigma$.

In order to quantify any effect that variations in the absolute flux
calibration of the near-infrared spectra have on our analysis, we have
absolutely flux calibrated each of the SpeX cross-dispersed spectra
separately using both 2MASS photometry and Mauna Kea Observatories
Near-Infrared \citep[MKO-NIR; ][]{2002PASP..114..180T} photometry
\citep{2004AJ....127.3553K}\footnote{The 2MASS $J$ band filter spans two
  spectral orders and therefore we have scaled these two orders to the
  same flux density levels by eye.  The two orders are then absolutely
  flux calibrated together using $J$-band photometry.}.  Two of the
dwarfs in our sample, 2MASS 1506$+$1321 and DENIS J0255$-$4700, have not
been observed in the MKO-NIR system to date.  The near-infrared, $L$
band, and IRS/SL spectra are then re-combined into a single 0.95$-$14.5
$\mu$m spectrum and the entire analysis is performed again.

Figure \ref{fig:FullErrorSum} summarizes the differences between the
atmospheric parameters derived by fitting the spectra with the different
absolute flux calibration schemes over the 0.95$-$14.5 $\mu$m wavelength
range.  Flux calibrating the spectra with 2MASS photometry leaves the
derived parameters almost unchanged while the derived parameters for
three objects flux calibrated with the MKO-NIR photometry change by 100
K in \teff\ and 0.5 dex in \logg.  Overall, there is little change in
the derived values of \teff, \logg, and \fsed\ which indicates the
absolute flux calibration does not severely affect the parameters
derived by fitting the 0.95$-$14.5 $\mu$m spectra (Figure
\ref{fig:YJHKErrorSum}).  The parameters obtained by fitting the
0.95$-$2.5 $\mu$m spectra alone are more sensitive to variations in the
absolute flux calibration.  Differences of 0.5$-$1 dex in \logg, 100 K
in \teff, and 1 in \fsed\ in the best fitting model spectra are
apparent.  The range of possible models (i.e., the error bars) also
grows because the SpeX cross-dispersed orders ($J$, $H$, $K$) are
absolutely flux calibrated separately and thus are varied separately in
the Monte Carlo simulation.

\subsection{Weights} \label{sec:Weights}

As discussed in \S\ref{sec:The Fitting Technique}, we weight each pixel
in the spectrum by its width ($w_i$ = $\Delta \lambda_i$) to remove a
bias introduced into the $G$ statistic due to the large variation in
wavelength sampling of the data.  Our choice of weights, however, is
completely arbitrary.  In order to estimate any systematic error in the
derived parameters introduced by our choice of weighting scheme, we have
devised a second weighting scheme as follows.

We divide the 0.95$-$14.5 $\mu$m wavelength range into bins
equally-spaced in $\ln \lambda$.  Each pixel that falls in bin $k$
receives a weight inversely proportional to the total number of pixels
contained in bin $k$. For example, if bin $k$ contains four pixels, then
$w_i$=0.25 for each pixel in bin $k$.  The \textit{Spitzer} data has the
lowest resolving power of all the data ($R \equiv \lambda / \Delta
\lambda \approx$ 100) so we used a bin size of $\delta \ln \lambda$ =
0.01 which corresponds to $R$=100. This alternative weighting scheme
effectively mimics smoothing the spectra to $R$=100 and resampling them
such that each pixel corresponds to a single resolution element.

Figure \ref{fig:WeightsFullErrorSum} summarizes the differences between
the atmospheric parameters derived by fitting the data using the two
different weight schemes over the 0.95$-$14.5 $\mu$m wavelength range.
The \teff\ values derived for the early-type L dwarfs are lower by 100 K
and the derived \logg\ values for two of the dwarfs change by 1 and 0.5.
Based on these results, the choice of weighting scheme does not
dramatically alter the results presented herein.

\section{Conclusions}

We have performed the first fits of the 0.95$-$14.5$\,\mu$m SEDs of nine
field ultracool dwarfs spanning spectral types L1 to T4.5 to obtain
effective temperatures, gravities and values for the cloud sedimentation
parameter.  The wide wavelength coverage of the data allows us to easily
discriminate between synthetic spectra with different atmospheric
parameters because most physical and chemical processes that occur in
ultracool dwarf atmospheres leave a signature in one or more of the
wavelength regions covered by the data.

In our study of field ultracool dwarfs we have focused on the most
significant physical parameters, ignoring the more subtle effects of
metallicity variations from the solar value as well as departures from
chemical equilibrium caused by vertical mixing in the atmosphere
(e.g. \citet{2006ApJ...647..552S}).  The spectroscopic data come from a
variety of instruments and we have estimated the uncertainties in the
derived parameters that arise from the absolute flux calibrations of the
various spectra that form the SEDs.  Although some wavelength ranges are
poorly fitted for some objects, undoubtedly due to remaining
inadequacies in the models and, to a smaller extent, to possible
variations in metallicity amongst the dwarfs in our sample, overall the
model spectra fit the data well.

Given the natural scatter in the observational properties among the
objects in our sample and the uncertainties inherent in our procedure,
we can identify only general trends in the fitted parameters as a
function of spectral type.  The effective temperatures decrease steadily
through the L1 to T4.5 spectral types and agree with the determinations
of \citet{2004AJ....127.3516G}. Our results also confirm that \teff\ is
nearly constant at the L/T transition, decreasing by only $\sim 200\,$K
from spectral types L7.5 to T4.5.

Two topics of great interest are the evolution of the cloud properties
as a function of spectral type and the distribution of cloud properties
within each spectral subclass.  Our work provides a first, although
somewhat blurry glimpse of this important aspect of ultracool dwarf
atmospheres.  We find that the cloud properties, as encapsulated in the
most important free parameter of the \citet{2001ApJ...556..872A}
condensation cloud model, \fsed, vary significantly among L dwarfs,
ranging from very thick clouds ($\fsed=1$) to relatively thin clouds
($\fsed=3$) with no particular trend with spectral type.  The two L
dwarfs in our sample with very red near-infrared colors have best
fitting model spectra with thick clouds.  Finally the fits to the two T
dwarfs in our sample (T2 and T4.5) suggest that the clouds becomes
thinner in this spectral class, in agreement with previous studies.

\begin{figure} 
  \centerline{\hbox{\includegraphics[scale=1.3]{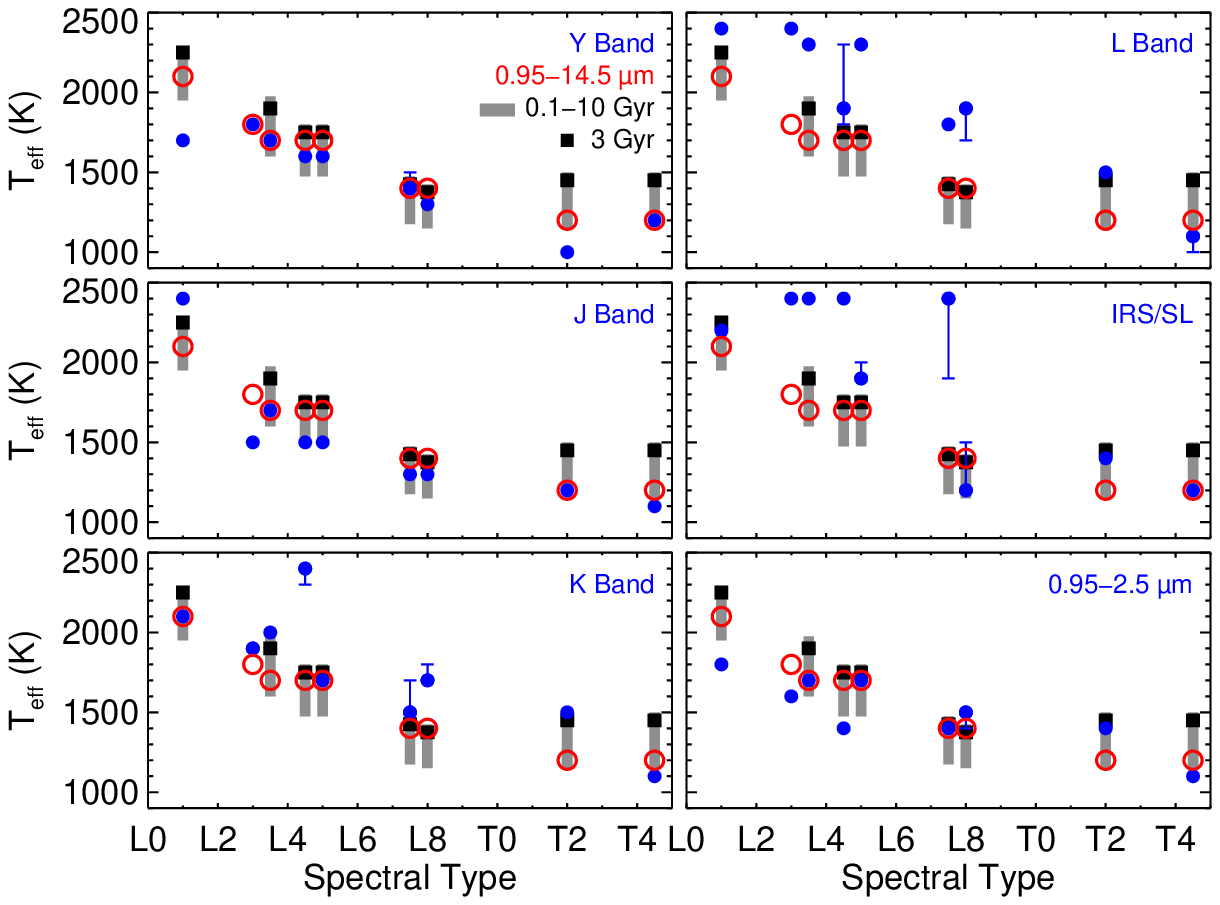}}}
  \caption{\label{fig:TeffSummary}Summary of the best fitting \teff\
    values (\textit{blue circles}) over the $Y$, $J$, $K$, and $L$ bands
    as well as the IRS/SL and the 0.95$-$2.5 $\mu$m wavelength ranges.
    Shown as open red circles are the best fitting \teff\ values derived
    by fitting the 0.95$-$14.5 $\mu$m spectra.  Errors bars denote the
    range of \teff\ values for the models with \fmc\ $>$ 0.1.  The grey
    regions denote the \teff\ range corresponding to an age range of
    0.1$-$10 Gyr and the black squares denote the \teff\ corresponding
    to 3 Gyr \citep{2004AJ....127.3516G}.}
\end{figure}

\begin{figure} 
\centerline{\hbox{\includegraphics[scale=1.3]{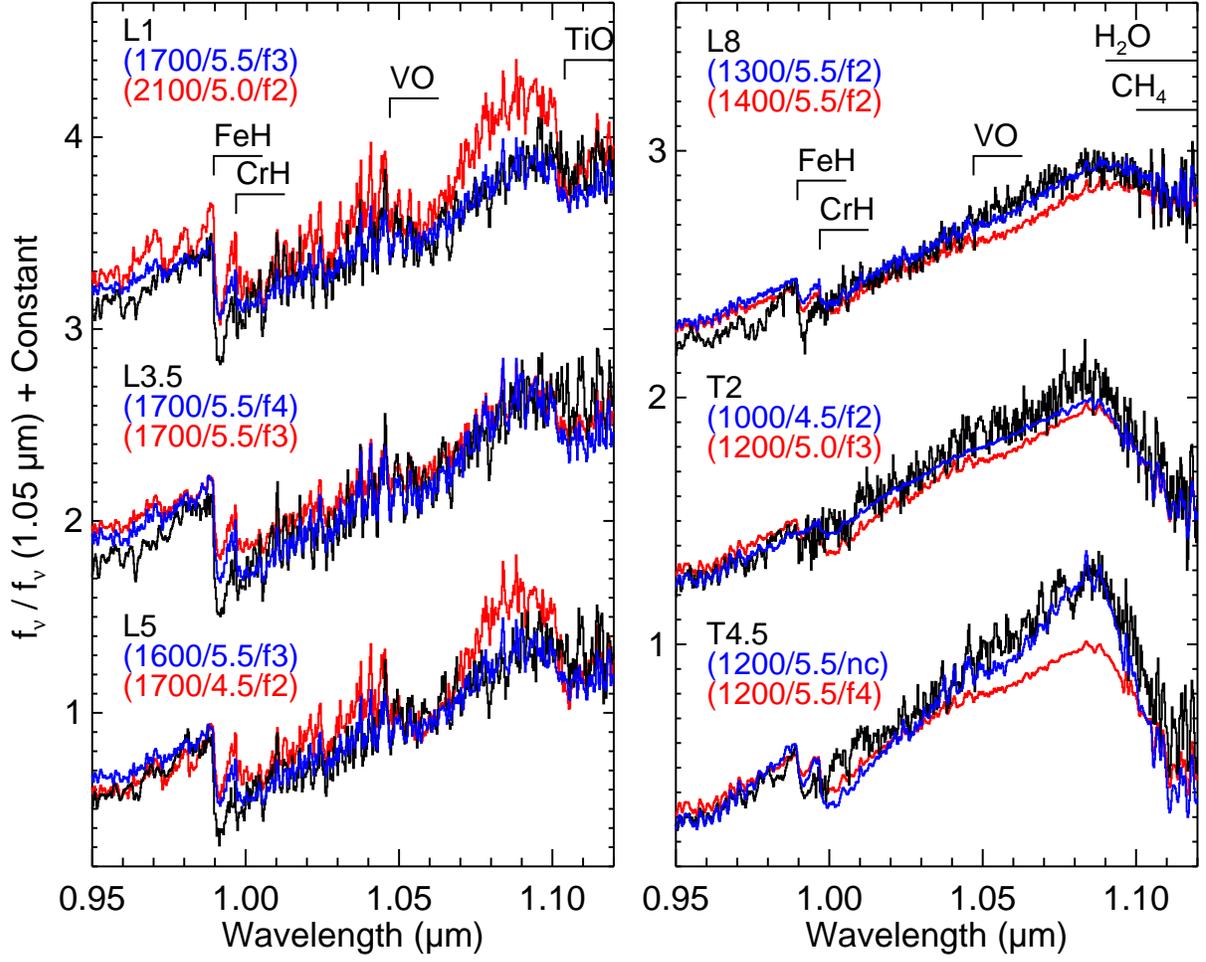}}}
\caption{\label{fig:YSeq2}The $Y$-band spectra of 2MASS 1439$+$1929
  (L1), 2MASS 0036$+$1821 (L3.5), 2MASS 1507$-$1627 (L5), DENIS
  0255$-$4700 (L8), SDSS 1254$-$0122 (T2), and 2MASS 0559$-$1404
  (T4.5) are shown in black.  The data have been normalized to unity at
  1.05 $\mu$m and offset for clarity.  The best fitting models
  (\teff/\logg/\fsed) are shown in blue as well as the best fitting
  models (\textit{red}) obtained from fitting the entire 0.95$-$14.5
  $\mu$m spectrum.  The models have been normalized and offset with the
  same constants as the data and were multiplied by the constant $C$
  (see \S\ref{sec:The Fitting Technique}) before normalization to
  preserve the relative flux levels between the data and models.}
\end{figure}

\begin{figure} 
\centerline{\hbox{\includegraphics[scale=1.3]{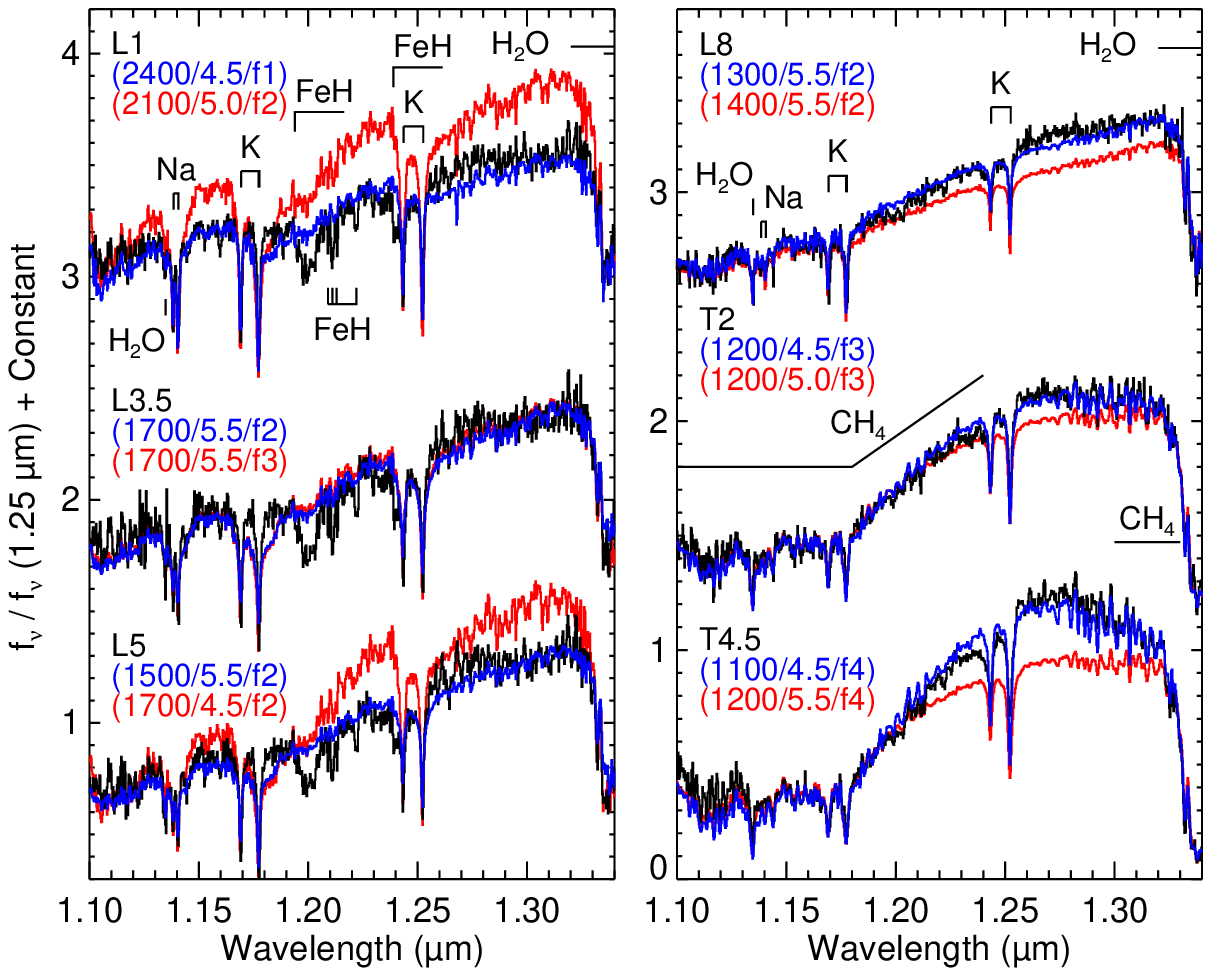}}}
\caption{\label{fig:JSeq2}Same as Figure \ref{fig:YSeq2} except the data
  cover the $J$ band and were normalized at 1.25 $\mu$m.}
\end{figure}

\begin{figure} 
\centerline{\hbox{\includegraphics[scale=1.3]{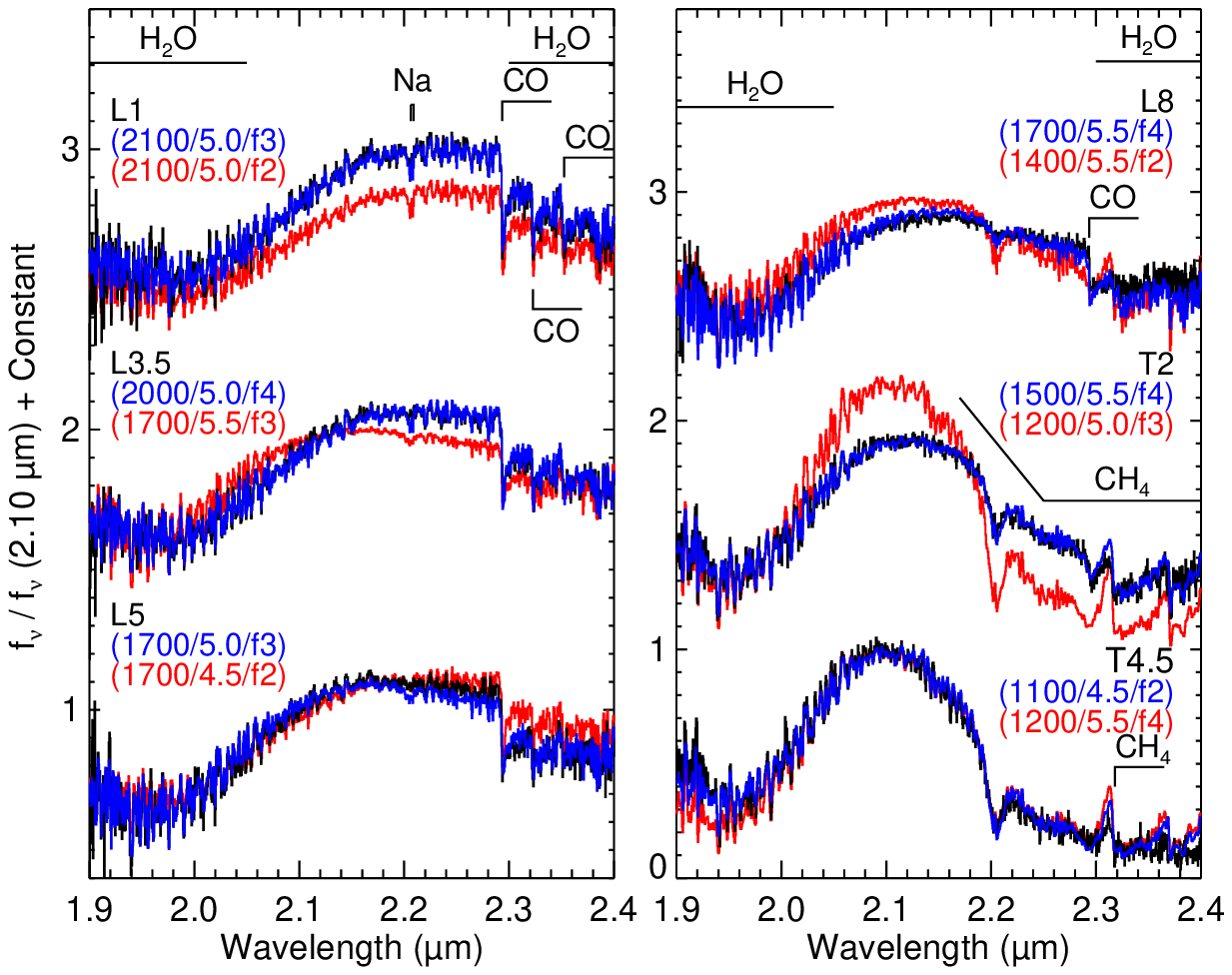}}}
\caption{\label{fig:KSeq2}Same as Figure \ref{fig:YSeq2} except the data
  cover the $K$ band and were normalized at 2.1 $\mu$m.}
\end{figure}

\begin{figure} 
\centerline{\hbox{\includegraphics[scale=1.5]{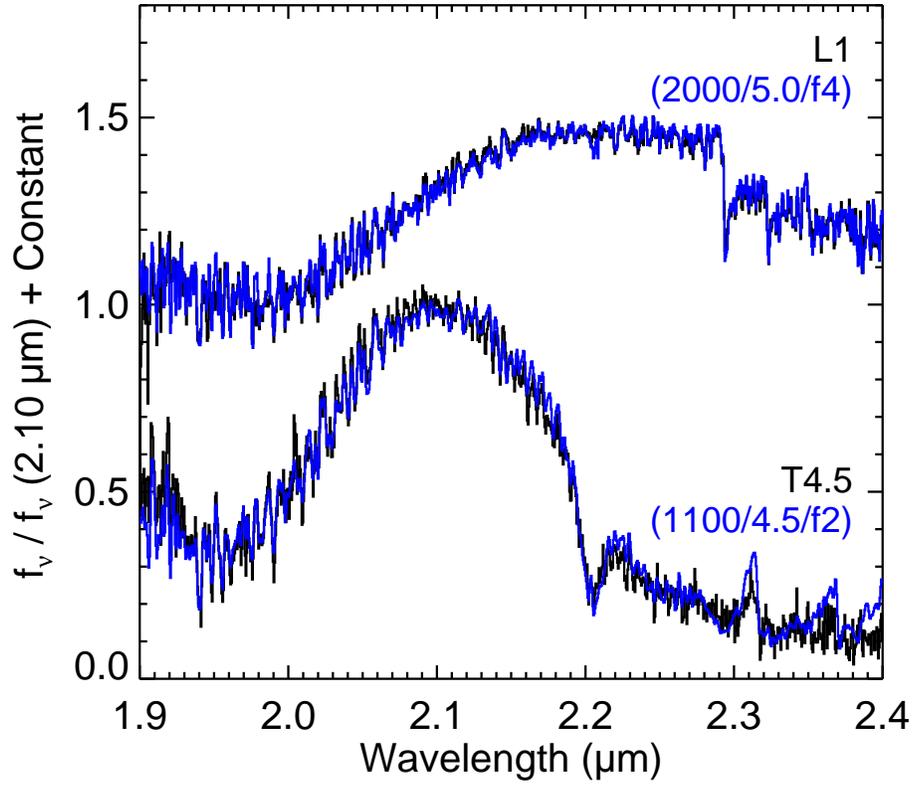}}}
\caption{\label{fig:KZoom}The $K$-band spectra of 2MASS 1439$+$1929 (L1)
  and and 2MASS 0559$-$1404 (T4.5) are shown in black.  The data have
  been normalized to unity at 2.1 $\mu$m and offset for clarity.  The
  best fitting models (\teff/\logg/\fsed) are shown in blue.  The models
  have been normalized and offset with the same constants as the data
  and were multiplied by the constant $C$ (see \S\ref{sec:The Fitting
    Technique}) before normalization to preserve the relative flux
  levels between the data and models.}
\end{figure}

\begin{figure} 
\centerline{\hbox{\includegraphics[scale=1.3]{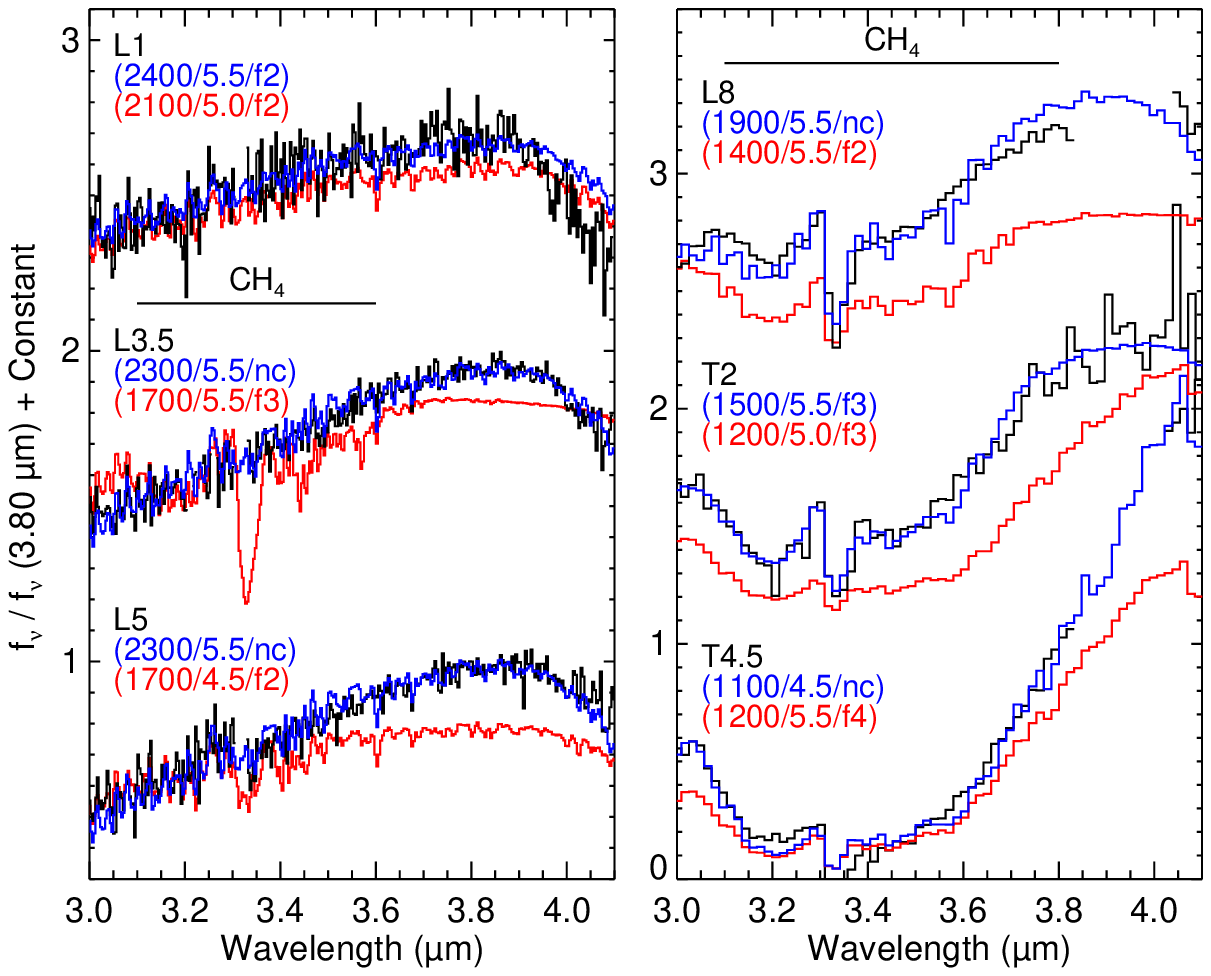}}}
\caption{\label{fig:LSeq2}Same as Figure \ref{fig:YSeq2} except the data
  cover the $L$ band and were normalized at 3.8 $\mu$m.}
\end{figure}

\begin{figure} 
\centerline{\hbox{\includegraphics[scale=1.3]{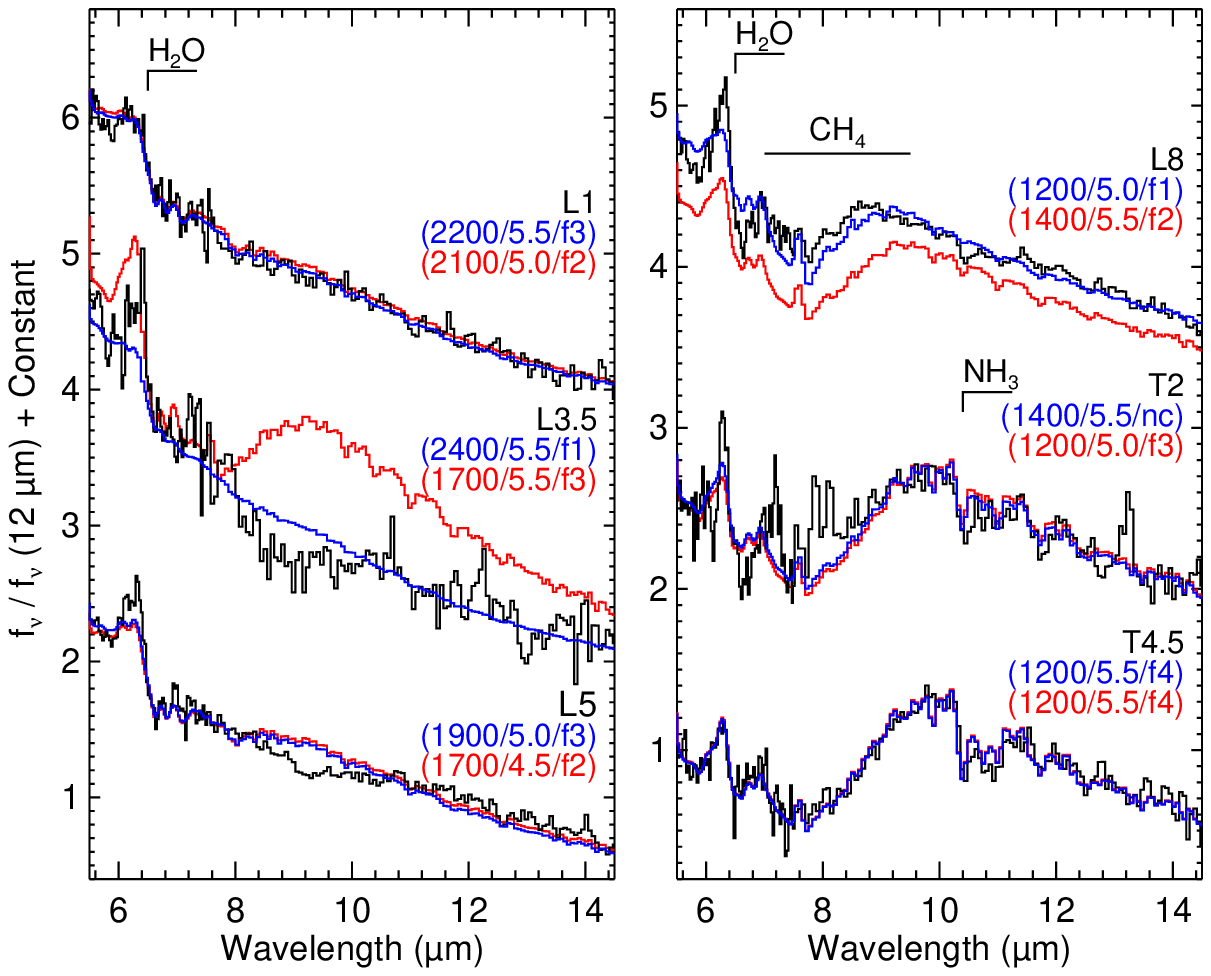}}}
\caption{\label{fig:IRSSeq2}Same as Figure \ref{fig:YSeq2} except the data
  cover the IRS/SL wavelength range and were normalized at 12 $\mu$m.}
\end{figure}

\begin{figure} 
\centerline{\hbox{\includegraphics[scale=1.3]{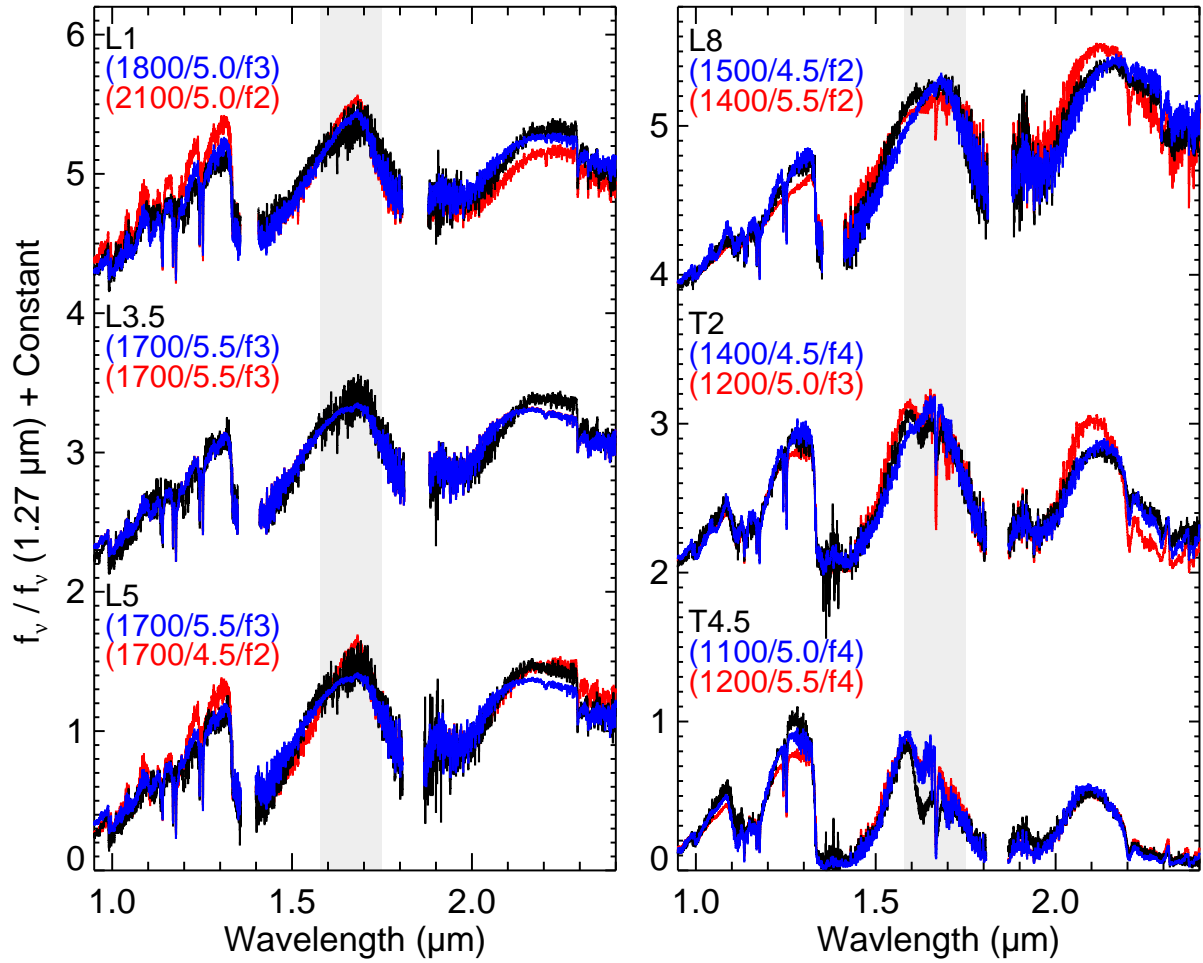}}}
\caption{\label{fig:YJHKSeq2}Same as Figure \ref{fig:YSeq2} except the
  data cover the 0.95$-$2.5 $\mu$m wavelength range and were normalized
  at 1.27 $\mu$m.}
\end{figure}

\begin{figure} 
\centerline{\hbox{\includegraphics[scale=1.5]{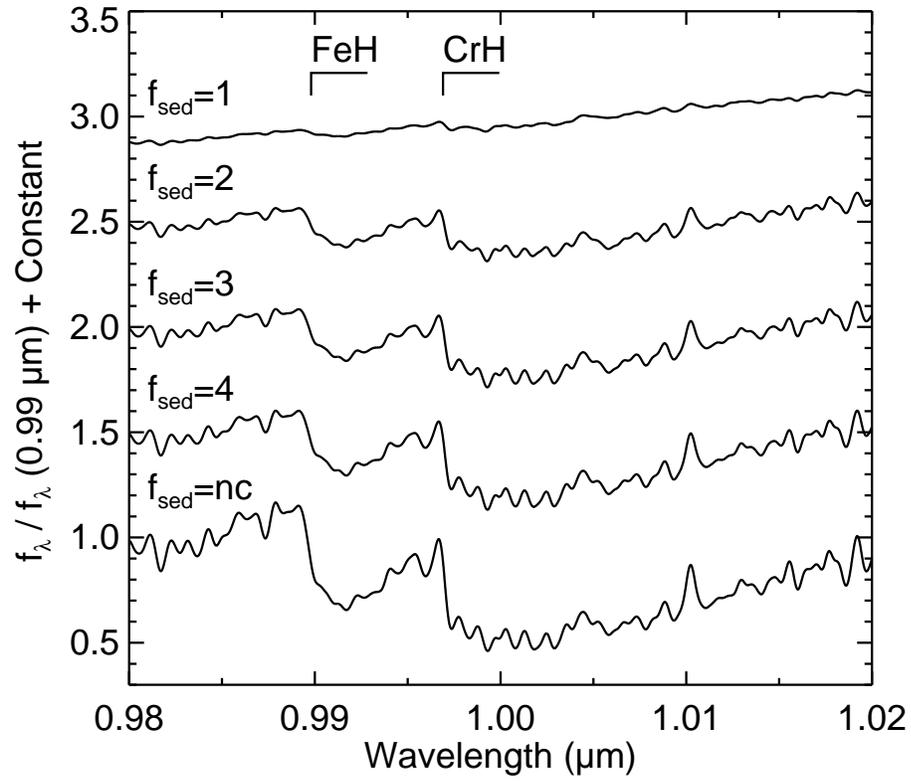}}}
\caption{\label{fig:FeH}Synthetic spectra smoothed to $R$=2000 with
  \teff=1400 K, \logg=5.0, and \fsed=1, 2, 3, 4, and the nc model.  The
  depth of the FeH bandhead at 0.9896 $\mu$m increases as \fsed\
  increases in value.}
\end{figure}

\begin{figure} 
\centerline{\hbox{\includegraphics[scale=1.4]{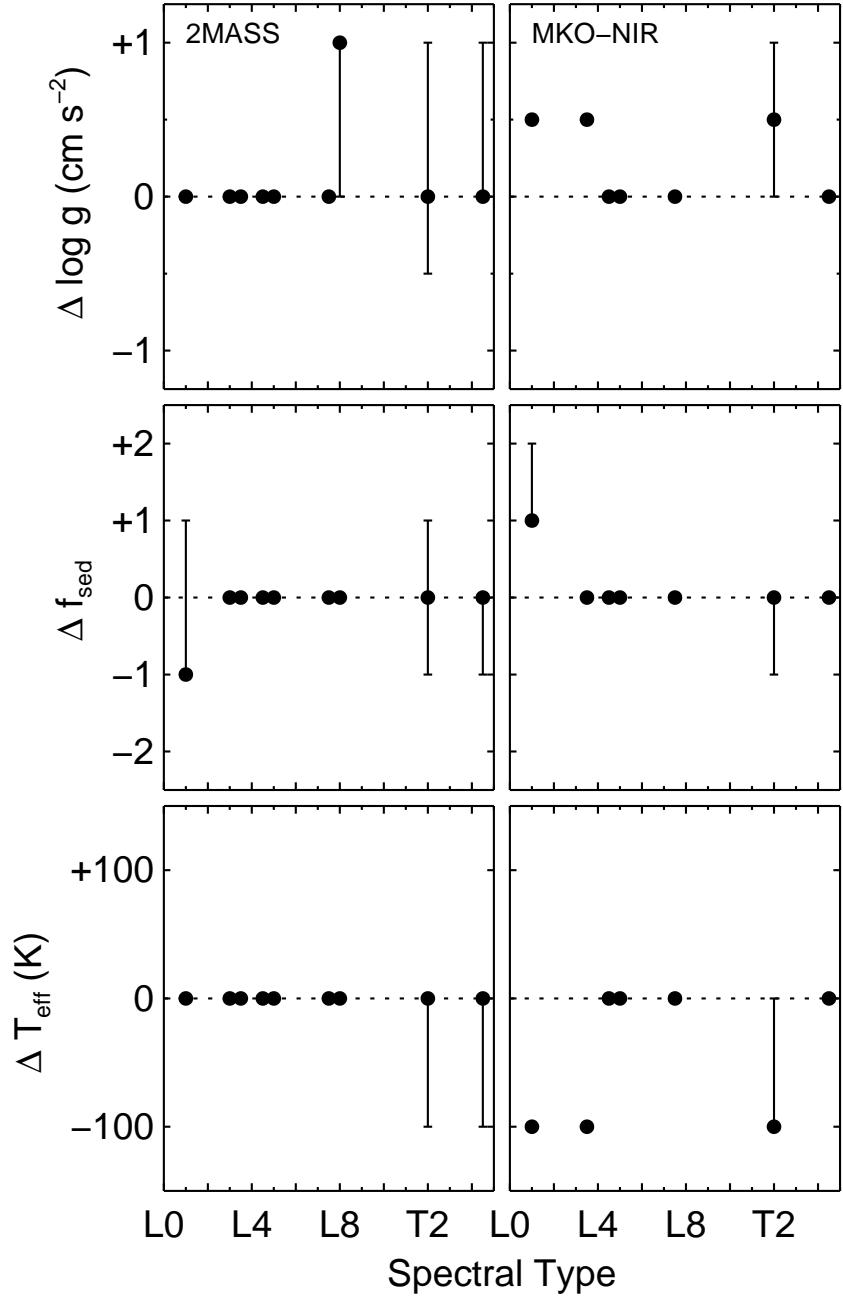}}}
\caption{\label{fig:FullErrorSum}Effects of different absolute flux
  calibrations on the derived atmospheric parameters over the
  0.95$-$14.5 $\mu$m wavelength range. The panels show the difference
  between the values derived after scaling the near-infrared spectra by
  a single scale factor and the values derived after scaling the $J$,
  $H$, and $K$ band spectra separately using 2MASS (\textit{left}) and
  MKO-NIR (\textit{right}) photometry.}
\end{figure}

\begin{figure} 
\centerline{\hbox{\includegraphics[scale=1.5]{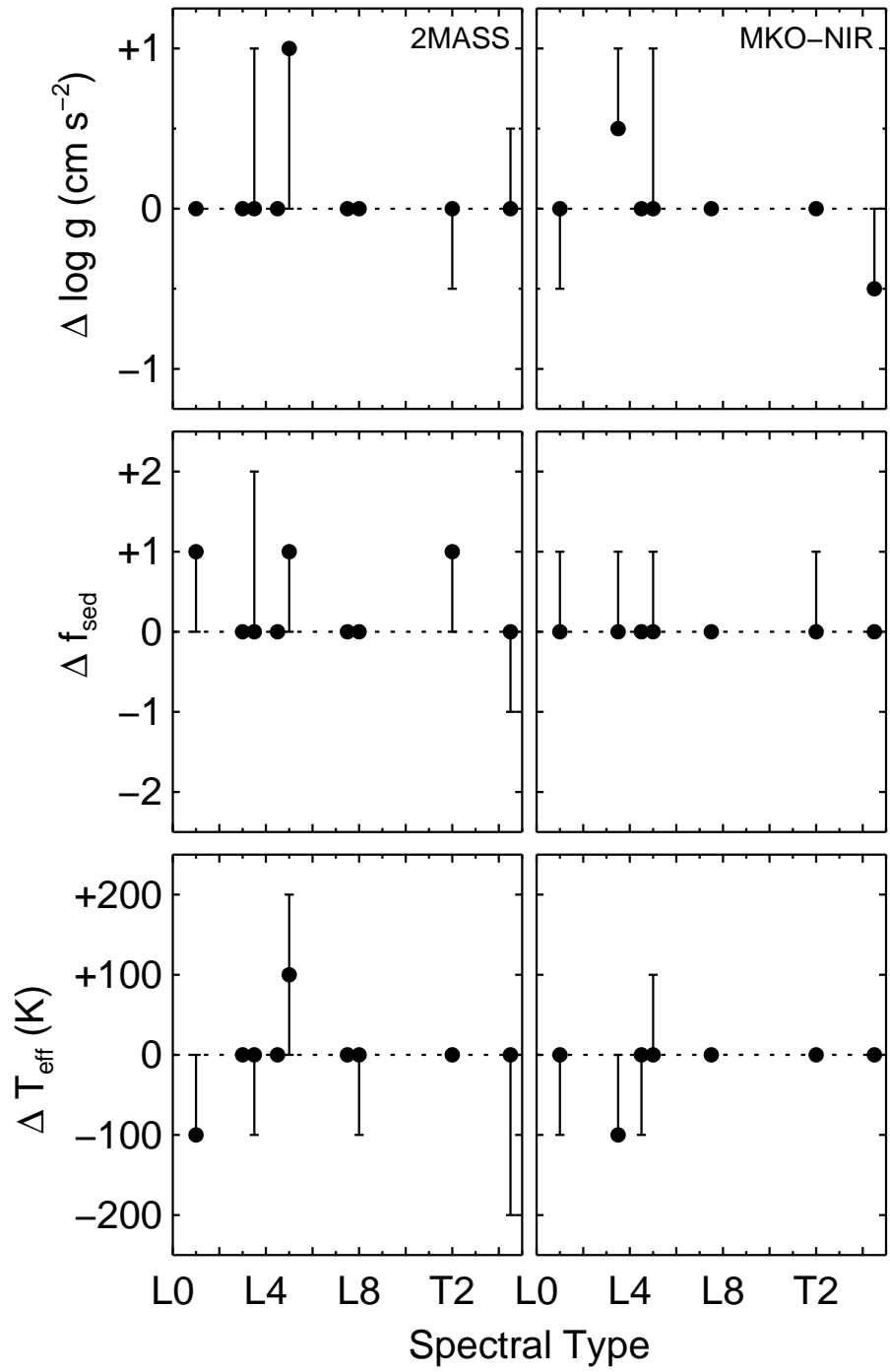}}}
\caption{\label{fig:YJHKErrorSum}Same as Figure \ref{fig:FullErrorSum}
  except the data were only fit from 0.95 to 2.5 $\mu$m.}
\end{figure}

\begin{figure} 
\centerline{\hbox{\includegraphics[scale=1.4]{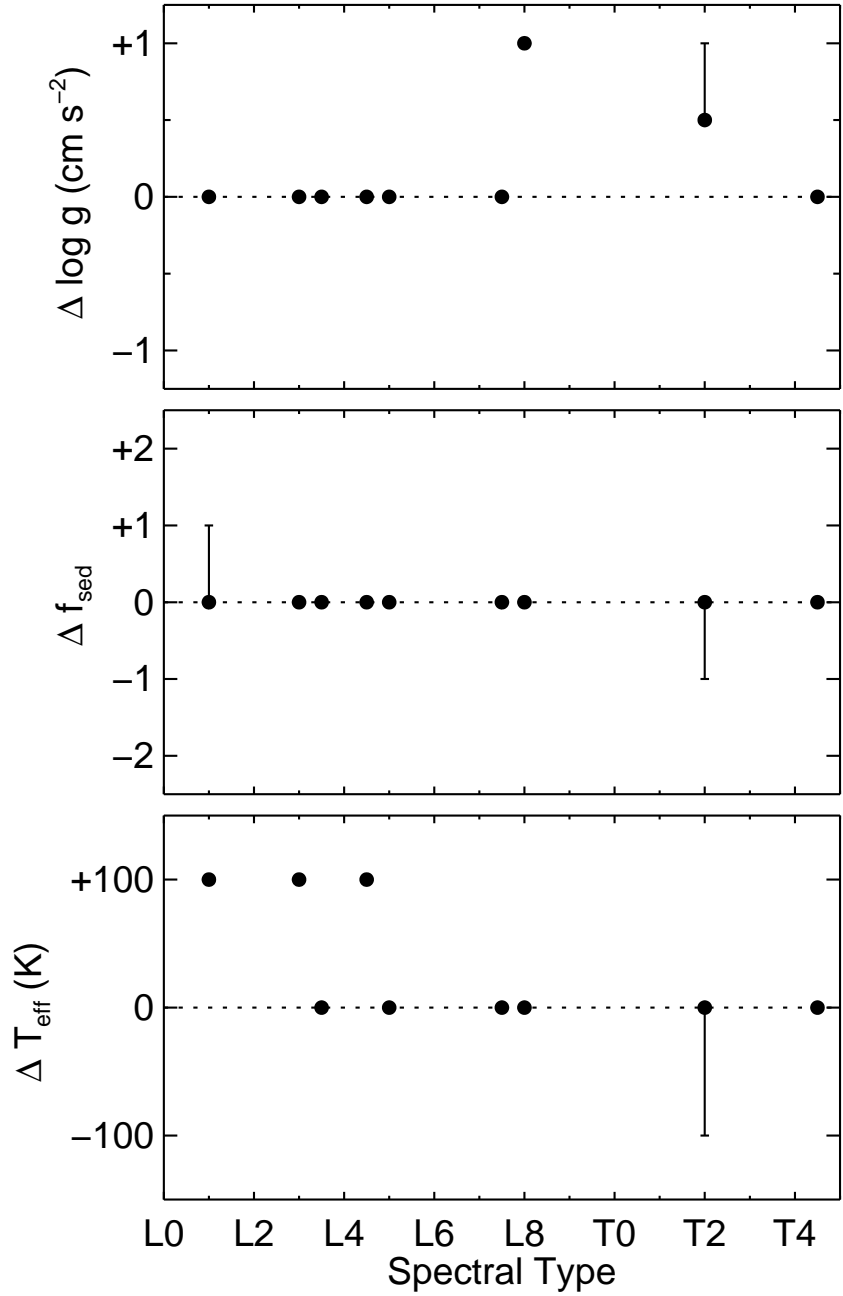}}}
\caption{\label{fig:WeightsFullErrorSum}Effects of different weighting
  schemes on the derived atmospheric parameters over the 0.95$-$14.5
  $\mu$m wavelength range.  The panels show the difference between the
  values derived by setting the weights to $w_i = \Delta \lambda_i$ and
  the values derived by setting the weights to $w_i = 1/n$ where $n$ is
  the number of pixels in bins of width $\delta \ln \lambda$=0.01.}
\end{figure}

Fitting individual spectral bands, we almost always obtain excellent
fits to the data, in some cases down to the fine structure of the H$_2$O
and CH$_4$ bands.  The \teff\ obtained in this fashion
(Fig. \ref{fig:TeffSummary}) can show quite a bit of scatter compared to
the values derived by fitting the full SED, typically by $\sim 200\,$K
and in the worst cases, up to 700$\,$K.  Nevertheless, the trend of
decreasing \teff\ with spectral type is preserved in all bands, with few
exceptions.  However, this exercise shows that in general, \teff\
determinations based on the analysis of narrow spectral ranges are not
reliable for ultracool dwarfs.

The data point to several directions for improving the models.  First,
improved input molecular opacities, particularly for CrH, VO, FeH, and
CH$_4$ and the resonance doublet of \ion{K}{1} (7665, 7699 \AA)
\citep{2007A&A...474L..21A} are required to improve fits to the early L
and all of the T dwarfs, respectively.  The second important area ripe
for improvement remains the cloud model, in particular the particles'
size distribution, the cloud's vertical profile, and the optical
properties of the condensates.

The accurate determination of the physical parameters of isolated field
ultracool dwarfs therefore remains an elusive goal.  Our method of SED
fitting should be applied to those ultracool dwarfs with well
constrained \teff\ ($\pm$20 K) and \logg\ ($\pm$0.05 dex) values
(derived from other methods) in order to better assess its internal
uncertainties.  Currently, such a sample is limited to a few late-type T
dwarfs
\citep{2006ApJ...647..552S,2007ApJ...656.1136S,2007ApJ...660.1507L} that
are companions to main sequence stars with known distances,
metallicities and well-constrained ages.  Such studies should be
extended to include earlier spectral types in order to firmly associate
the observable properties of the L and T spectral sequence to
fundamental astrophysical parameters.

\acknowledgements

We thank Adam Burgasser and Kelle Cruz for useful discussions, Mike Liu
for suggestions that improved the original manuscript, and the anonymous
referee for his/her careful review of the manuscipt and constructive
comments.  This publication makes use of data from the Two Micron All
Sky Survey, which is a joint project of the University of Massachusetts
and the Infrared Processing and Analysis Center, and funded by the
National Aeronautics and Space Administration and the National Science
Foundation, the SIMBAD database, operated at CDS, Strasbourg, France,
NASA's Astrophysics Data System Bibliographic Services, the M, L, and T
dwarf compendium housed at DwarfArchives.org and maintained by Chris
Gelino, Davy Kirkpatrick, and Adam Burgasser, and the NASA/ IPAC
Infrared Science Archive, which is operated by the Jet Propulsion
Laboratory, California Institute of Technology, under contract with the
National Aeronautics and Space Administration.  This work is based (in
part) on observations made with the \textit{Spitzer Space Telescope},
which is operated by the Jet Propulsion Laboratory, California Institute
of Technology under a contract with NASA and is supported (in part) by
the United States Department of Energy under contract W-7405-ENG-36,
NASA through the Spitzer Space Telescope Fellowship Program, through a
contract issued by the Jet Propulsion Laboratory, California Institute
of Technology under a contract with NASA.  Work by K.L. is supported by
NSF grant AST0406963, NASA grant NNG06GC26G, and the NASA Office of
Space Science.  M.S.M acknowledges support from the NASA Office of Space
Science.

{\it Facilities:} \facility{Keck (LRIS)}, \facility{IRTF (SpeX)}, \facility{Subaru (IRCS), \facility{Spitzer (IRS)}}.

\clearpage

\bibliographystyle{apj}
\bibliography{ref,tmp}

\clearpage 

\clearpage 

\clearpage 

\clearpage 

\clearpage

\clearpage

\clearpage

\clearpage

\clearpage

\clearpage

\clearpage

\clearpage

\clearpage

\clearpage

\clearpage

\clearpage

\clearpage

\clearpage

\clearpage

\clearpage

\clearpage

\clearpage

\clearpage

\clearpage

\clearpage

\input{tab1.tex}

\clearpage

\input{tab2.tex}

\clearpage

\input{tab3.tex}

\clearpage

\input{tab4.tex}

\clearpage
\begin{landscape}
\input{tab5.tex}
\clearpage
\end{landscape}

\end{document}

%% file: tab1.tex
\begin{deluxetable}{lllcccccl}
\tablecolumns{7}
\tabletypesize{\scriptsize} 
\tablewidth{0pc}
\tablecaption{\label{tab:objlist}The Properties of the L and T Dwarfs in Our Sample}
\tablehead{
\colhead{Object} & 
\multicolumn{2}{c}{Spectral Type\tablenotemark{a}} & 
\colhead{$J$\tablenotemark{b}} & 
\colhead{$J-H$} &
\colhead{$H-K_s$} &
\colhead{$J-K_s$} & 
\colhead{$\pi$\tablenotemark{c}} & 
\colhead{Ref.} \\ 
\cline{2-3}
\colhead{} & 
\colhead{Optical} &
\colhead{Infrared} &
\colhead{(mag)} & 
\colhead{(mag)} &
\colhead{(mag)} &
\colhead{(mag)} & 
\colhead{(mas)} & 
\colhead{} }

\startdata

2MASS J14392836$+$1929149   & L1       & L1         &  12.759$\pm$0.019 & 0.718$\pm$0.027 & 0.495$\pm$0.029 & 1.213$\pm$0.029 & \phn69.6$\pm$0.5  & 1 \\
2MASS J15065441$+$1321060   & L3       & $\cdots$   &  13.365$\pm$0.023 & 0.985$\pm$0.031 & 0.639$\pm$0.028 & 1.624$\pm$0.030 & $\cdots$          & $\cdots$ \\
2MASS J00361617$+$1821104   & L3.5   & L4$\pm$1     &  12.466$\pm$0.027 & 0.878$\pm$0.040 & 0.530$\pm$0.036 & 1.408$\pm$0.034 & 114.2$\pm$0.8     & 1 \\
2MASS J22244381$-$0158521   & L4.5     & L3.5       &  14.073$\pm$0.027 & 1.255$\pm$0.037 & 0.796$\pm$0.035 & 2.051$\pm$0.035 & \phn87.02$\pm$0.89  & 1,2 \\
2MASS J15074769$-$1627386   & L5       & L5.5       &  12.830$\pm$0.027 & 0.935$\pm$0.036 & 0.583$\pm$0.035 & 1.518$\pm$0.037 & 136.4$\pm$0.6     & 1 \\
2MASS J08251968$+$2115521   & L7.5     & L6         &  15.100$\pm$0.034 & 1.308$\pm$0.047 & 0.764$\pm$0.041 & 2.072$\pm$0.043 & \phn94.22$\pm$0.99  & 1,2 \\
DENIS 025503.3$-$470049.0   & L8       & L9         &  13.246$\pm$0.027 & 1.042$\pm$0.036 & 0.646$\pm$0.034 & 1.688$\pm$0.036 & 201.37$\pm$3.89   & 3 \\
SDSS J125453.90$-$012247.4  & T2       & T2         &  14.891$\pm$0.035 & 0.801$\pm$0.043 & 0.253$\pm$0.060 & 1.054$\pm$0.064 & \phn73.96$\pm$1.59  & 2,4 \\
2MASS J05591914$-$1404488   & T5       & T4.5       &  13.802$\pm$0.024 & 0.123$\pm$0.050 & 0.102$\pm$0.068 & 0.225$\pm$0.057 & \phn96.73$\pm$0.96  & 1,2 \\

\enddata
\tablerefs{(1) \citet{2002AJ....124.1170D}, (2) \citet{2004AJ....127.2948V}, (3) \citet{2006AJ....132.1234C}, and (4) \citet{2003AJ....126..975T}}

\tablenotetext{a}{Spectral types of the L dwarfs are from
  \citet{1999ApJ...519..802K,2000AJ....120..447K},
  \citet{2000AJ....119..369R}, \citet{2000AJ....120.1085G},
  \citet{2002ApJ...564..466G}, \citet{2004AJ....127.3553K},
  \citet{2006ApJ...637.1067B}, and J.D.  Kirkpatrick (in preparation).
  Spectral types of the T dwarfs are from \citet{2006ApJ...637.1067B}.
  Errors on spectral types are $\pm$0.5 subclass unless otherwise
  noted.}

\tablenotetext{b}{$J$-, $H$-, and $K_s$-band photometry is from the 2MASS Point Source Catalog.}

\tablenotetext{c}{Based on the weighted mean of the referenced parallaxes \citep{2004AJ....127.3516G}}

\end{deluxetable}

%% file: tab2.tex
\begin{deluxetable}{lcccccl}
\tablecolumns{7}
\tabletypesize{\scriptsize} 
\tablewidth{0pc}
\tablecaption{\label{tab:specinfo}Spectral Characteristics of Ultracool Dwarfs in Our Sample}
\tablehead{
\colhead{Object} & 
\colhead{} & 
\multicolumn{4}{c}{$R$=$\lambda / \Delta \lambda$} & 
\colhead{Ref.} \\ 
\cline{2-6}
\colhead{} & 
\colhead{} & 
\colhead{(0.6$-$0.9 $\mu$m)} & 
\colhead{(0.9$-$2.5 $\mu$m)} &
\colhead{(3.0$-$4.1 $\mu$m)} &
\colhead{(5.5$-$14.5 $\mu$m)} & 
\colhead{}}

\startdata

2MASS J1439$+$1929   & & 890 & 2000 & 940 & 90 & 1,2,2,3 \\
2MASS J1506$+$1321   & & 890 & 2000 & 940 & 90 & 4,2,2,3 \\
2MASS J0036$+$1821   & & 890 & 2000 & 940 & 90 & 5,2,2,3 \\
2MASS J2224$-$0158   & & 890 & 2000 & 425 & 90 & 6,2,2,3 \\
2MASS J1507$-$1627   & & 890 & 2000 & 940 & 90 & 5,2,2,3 \\
2MASS J0825$+$2115   & & 890 & 1200 & 210 & 90 & 6,2,2,3 \\
DENIS 0255$-$4700    & & 890 & 2000 & 210 & 90 & 7,2,2,3 \\
SDSS J1254$-$0122    & & 1200 & 1200 & 210 & 90 & 8,2,2,3 \\
2MASS J0559$-$1404   & & 1200 & 1200 & 210 & 90 & 8,2,2,3 \\

\enddata

\tablerefs{(1) \citet{1999ApJ...519..802K}; (2)
  \citet{2005ApJ...623.1115C}; (3) \citet{2006ApJ...648..614C}; (4)
  \citet{2000AJ....120.1085G}; (5) \citet{2000AJ....119..369R}; (6)
  \citet{2000AJ....120..447K}; (7) Kirkpatrick (2007, private
  communication); (8) \citet{2003ApJ...594..510B} }

\end{deluxetable}

%% file: tab3.tex
\begin{deluxetable}{lllccccccccc}
\tablecolumns{12} 
\tabletypesize{\scriptsize} \tablewidth{0pc}
\tablecaption{\label{tab:Full}Derived Parameters for L and T Dwarfs
    (0.9$-$14.5 $\mu$m)} 
\tablehead{
\colhead{Object} & 
\multicolumn{2}{c}{} & 
\colhead{} &  
\multicolumn{5}{c}{Atmospheric Parameters} & 
\multicolumn{3}{c}{} \\ 
\cline{5-9}
\colhead{} & 
\multicolumn{2}{c}{Spectral Type\tablenotemark{a}} &  
\colhead{} &  
\multicolumn{4}{c}{By Spectral Fitting} & 
\colhead{By Evolutionary Sequences} & 
\multicolumn{3}{c}{Golimowski et al. $T_{\mathrm{eff}}$ (K)} \\
\cline{2-3}
\cline{5-8}
\cline{10-12} 
\colhead{} &
\colhead{Optical} & 
\colhead{Infrared} &
\colhead{} &  
\colhead{$T_{\mathrm{eff}}$ (K)} & 
\colhead{$\log g$ (cm s$^{-2}$)} &
\colhead{$f_{\mathrm{sed}}$} &
\colhead{$f_{\mathrm{MC}}$} &
\colhead{$\log g$ (cm s$^{-2}$)} &
\colhead{0.1 Gyr} & 
\colhead{3 Gyr} & 
\colhead{10 Gyr}} 

\startdata
 
2MASS J1439$+$1929                 & L1   & L1       & & 2100 & 5.0 &  2 & 0.582 & 5.188    & 1950 & 2250 & 2275  \\
                                   &      &          & & 2100 & 5.0 &  3 & 0.418 & 5.188    & 1950 & 2250 & 2275 \\
2MASS J1506$+$1321                 & L3   & $\cdots$ & & 1800 & 4.5 &  1 & 1.000 & $\cdots$ & $\cdots$ & $\cdots$ & $\cdots$ \\
2MASS J0036$+$1821                 & L3.5 & L4$\pm$1 & & 1700 & 5.5 &  3 & 1.000 & 4.924    & 1650 & 1900 & 1975 \\
2MASS J2224$-$0158                 & L4.5 & L3.5     & & 1700 & 4.5 &  1 & 0.999 & 5.378    & 1475 & 1750 & 1800 \\
2MASS J1507$-$1627                 & L5   & L5.5     & & 1700 & 4.5 &  2 & 1.000 & 5.451    & 1475 & 1750 & 1800  \\
2MASS J0825$+$2115                 & L7.5 & L6       & & 1400 & 4.5 &  1 & 1.000 & 5.482    & 1175 & 1425 & 1475 \\
DENIS 0255$-$4700                  & L8   & L9       & & 1400 & 5.5 &  2 & 0.988 & 5.547    & 1150\tablenotemark{b} & 1375 & 1425 \\
SDSS J1254$-$0122                  & T2   & T2       & & 1200 & 5.0 &  3 & 0.831 & 4.799    & 1150 & 1425 & 1500 \\
                                   &      &          & & 1200 & 5.5 &  2 & 0.105 & 4.908    & 1150 & 1425 & 1500  \\
2MASS J0559$-$1404                 & T5   & T4.5     & & 1200 & 5.5 &  4 & 0.999 & 4.713    & 1150 & 1425 & 1500  \\
 
\enddata
 
\tablecomments{Only models with $f_{\mathrm{MC}}$ $>$ 0.1 are listed.}
\tablenotetext{a}{Spectral types of the L dwarfs are from
  \citet{1999ApJ...519..802K}, \citet{2000AJ....119..369R},
  \citet{2000AJ....120.1085G}, \citet{2000AJ....120..447K},
  \citet{2006ApJ...637.1067B}, and J.D.  Kirkpatrick (in preparation).
  Spectral types of the T dwarfs are from \citet{2003ApJ...594..510B,2006ApJ...637.1067B}. 
  Errors on spectral types are
  $\pm$0.5 subclass unless otherwise noted.}

\tablenotetext{b}{DENIS 0255$-$4700 was not included in the
  \citet{2004AJ....127.3516G} sample.  We have computed its
  $L_{\mathrm{bol}}$ using the new parallax measurement of
  \citet{2006AJ....132.1234C}.  The bolometric flux was determined by
  integrating over its 0.6 to 14.5 $\mu$m spectrum as described in
  \citet{2006ApJ...648..614C}.}

\end{deluxetable}

%% file: tab4.tex
\begin{deluxetable}{lccccc}
\tablecolumns{6}
\tabletypesize{\scriptsize} 
\tablewidth{0pc}
\tablecaption{\label{tab:BinaryTable}The Properties of SDSS 1254$-$0122 and 2MASS 0559$-$1404}
\tablehead{
\colhead{} & 
\multicolumn{2}{c} {{SDSS 1254$-$0122}} & 
\colhead{} & 
\multicolumn{2}{c} {{2MASS 0559$-$1404}}  \\
\cline{2-3}
\cline{5-6}
\colhead{Parameter} & 
\colhead{single} & 
\colhead{equal mass binary} &
\colhead{} & 
\colhead{single} &
\colhead{equal mass binary}}

\startdata

$\log$ ($L_{\mathrm{bol}}$/$L_{\mathrm{\odot}}$) &  $-$4.60   & $-$4.90 & & $-$4.63   & $-$4.93 \\
$T_{\mathrm{eff}}$ (K)                           &  1200      & $<$1160 & & 1200      & $<$1150 \\
$\log g$ (cm s$^{-2}$)                           &  4.71      & $<$5.38 & & 4.80      & $<$5.38 \\
Age (Gyr)                                        &  $\sim$0.3 & $<$10   & & $\sim$0.4 & $<$ 10 \\

\enddata

\end{deluxetable}

%% file: tab5.tex
\begin{deluxetable}{lllccccccc}
\tablecolumns{10} 
\tabletypesize{\scriptsize} \tablewidth{0pc}
\tablecaption{\label{tab:MicroFits}Derived Parameters of the L and T Dwarfs from Restricted Spectral Intervals} 
\tablehead{
\colhead{Object} & 
\multicolumn{2}{c} {Spectral Type\tablenotemark{a}} &  
\colhead{} &  
\multicolumn{6}{c} {\teff (K) / \logg (cm s$^{-2}$) / \fsed\ /  \fmc} \\
\cline{2-3}
\cline{5-10}
 \colhead{} &
\colhead{Optical} & 
\colhead{Infrared} & 
\colhead{} &  
\colhead{$Y$} &
\colhead{$J$} & 
\colhead{$K$} &  
\colhead{$L$} &  
\colhead{IRS/SL} &  
\colhead{Near-Infrared}}
 
\startdata
 
2MASS 1439$+$1929 & L1   & L1       &  & 1700/5.5/3/1.000 & 2400/4.5/1/1.000 & 2100/5.0/3/1.000 & 2400/5.5/2/0.922 & 2200/5.5/3/0.922 & 1800/5.0/3/1.000 \\
2MASS 1506$+$1321 & L3   & $\cdots$ &  & 1800/5.5/4/1.000 & 1500/5.5/1/1.000 & 1900/4.5/3/0.761 & 2400/5.5/4/0.997 & 2400/5.5/1/1.000 & 1600/4.5/2/1.000 \\
                  &      &          &  & $\cdots$          & $\cdots$          & 1900/4.5/2/0.240 & $\cdots$          & $\cdots$          & $\cdots$          \\
2MASS 0036$+$1821 & L3.5 & L4$\pm$1 &  & 1700/5.5/4/0.988 & 1700/5.5/2/1.000 & 2000/5.0/4/1.000 & 2300/5.5/nc/1.000 & 2400/5.5/1/0.863 & 1700/5.5/3/1.000 \\
2MASS 2224$-$0158 & L4.5 & L3.5     &  & 1600/5.5/3/0.996 & 1500/5.0/1/1.000 & 2400/5.5/nc/0.748 & 1900/4.5/4/0.219 & 2400/5.5/1/1.000 & 1400/4.5/1/1.000 \\
                  &      &          &  & $\cdots$          & $\cdots$          & 2300/5.0/nc/0.133 & 2200/5.5/nc/0.190 & $\cdots$          & $\cdots$          \\
                  &      &          &  & $\cdots$          & $\cdots$          & $\cdots$          & 2300/5.5/nc/0.122 & $\cdots$          & $\cdots$          \\
                  &      &          &  & $\cdots$          & $\cdots$          & $\cdots$          & 1800/5.0/4/0.114 & $\cdots$          & $\cdots$          \\
2MASS 1507$-$1627 & L5   & L5.5     &  & 1600/5.5/3/1.000 & 1500/5.5/2/0.999 & 1700/5.0/3/1.000 & 2300/5.5/nc/1.000 & 1900/5.0/3/0.627 & 1700/5.5/3/1.000 \\
                  &      &          &  & $\cdots$          & $\cdots$          & $\cdots$          & $\cdots$          & 2000/5.0/3/0.370 & $\cdots$          \\
2MASS 0825$+$2115 & L7.5 & L6       &  & 1400/5.5/2/0.549 & 1300/4.5/1/0.984 & 1500/4.5/2/0.835 & 1800/5.0/4/0.998 & 2400/5.5/nc/0.201 & 1400/4.5/1/1.000 \\
                  &      &          &  & 1500/5.0/2/0.359 & $\cdots$          & 1700/5.0/3/0.166 & $\cdots$          & 2200/5.5/4/0.166 & $\cdots$          \\
                  &      &          &  & $\cdots$          & $\cdots$          & $\cdots$          & $\cdots$          & 2300/5.5/nc/0.157 & $\cdots$          \\
                  &      &          &  & $\cdots$          & $\cdots$          & $\cdots$          & $\cdots$          & 1900/5.5/4/0.142 & $\cdots$          \\
DENIS 0255$-$4700 & L8   & L9       &  & 1300/5.5/2/0.992 & 1300/5.5/2/0.999 & 1700/5.5/4/0.820 & 1900/5.5/nc/0.447 & 1200/5.5/1/0.465 & 1500/4.5/2/0.572 \\
                  &      &          &  & $\cdots$          & $\cdots$          & 1800/5.5/4/0.177 & 1700/5.5/4/0.241 & 1500/4.5/2/0.218 & 1400/4.5/2/0.428 \\
                  &      &          &  & $\cdots$          & $\cdots$          & $\cdots$          & 1700/5.0/4/0.186 & 1200/5.0/1/0.167 & $\cdots$          \\
SDSS 1254$-$0122  & T2   & T2       &  & 1000/4.5/2/1.000 & 1200/4.5/3/1.000 & 1500/5.5/4/1.000 & 1500/5.5/3/0.918 & 1400/5.5/nc/0.910 & 1400/4.5/4/1.000 \\
2MASS 0559$-$1404 & T4.5 & T4.5     &  & 1200/5.5/nc/1.000 & 1100/4.5/4/1.000 & 1100/4.5/2/1.000 & 1100/4.5/nc/0.469 & 1200/5.5/4/0.982 & 1100/5.0/4/1.000 \\
                  &      &          &  & $\cdots$          & $\cdots$          & $\cdots$          & 1000/4.5/3/0.366 & $\cdots$          & $\cdots$          \\
\enddata
 
\tablecomments{Only models with $f_{\mathrm{MC}}$ $>$ 0.1 are listed.}
\tablenotetext{a}{Spectral types of the L dwarfs are from
  \citet{1999ApJ...519..802K}, \citet{2000AJ....119..369R},
  \citet{2000AJ....120.1085G}, \citet{2000AJ....120..447K},
  \citet{2006ApJ...637.1067B}, and J.D.  Kirkpatrick (in preparation).
  Spectral types of the T dwarfs are from \citet{2003ApJ...594..510B,2006ApJ...637.1067B}.  
  Errors on spectral types are
  $\pm$0.5 subclass unless otherwise noted.}
 
\end{deluxetable}